\documentclass[preprint,12pt]{elsarticle}

\usepackage{graphicx}

\usepackage{amsfonts,amstext,amsmath,amssymb,amsthm}

\usepackage{rotating}

\usepackage{empheq}

\usepackage{times}
\usepackage{subcaption} 
\usepackage{wrapfig}

\usepackage{makecell}
\usepackage{tabularx, booktabs}
\newcolumntype{Y}{>{\centering\arraybackslash}X}

\usepackage{multirow}
\usepackage{bigstrut}

\usepackage{paralist}

\usepackage{setspace}

\usepackage[%
    bookmarks=false,         
    unicode=false,          
    pdftoolbar=true,        
    pdfmenubar=true,        
    pdffitwindow=false,     
    pdfstartview={FitH},    
    pdftitle={On Robustness of the Shiryaev--Roberts Procedure for Quickest Change-Point Detection under Parameter Misspecification in the Post-Change Distribution},    
    pdfauthor={Wenyu Du and Aleksey S. Polunchenko and Grigory Sokolov},     
    pdfsubject={On Robustness of the Shiryaev--Roberts Procedure for Quickest Change-Point Detection under Parameter Misspecification in the Post-Change Distribution},   
    pdfcreator={Aleksey S. Polunchenko},   
    pdfproducer={MikTeX}, 
    pdfkeywords={Sequential Analysis}{Quickest change-point detection}{Shiryaev--Roberts procedure}{Robustness analysis}, 
    pdfnewwindow=true,      
    colorlinks=false,       
    linkcolor=red,          
    citecolor=green,        
    filecolor=magenta,      
    urlcolor=cyan           
]{hyperref}

\usepackage[utf8]{inputenc}
\usepackage[T1]{fontenc}
\usepackage{textcomp}

\biboptions{authoryear,compress,round}

\newcommand{\ignore}[1]{}

\renewcommand{\Pr}{\mathbb{P}} 
\DeclareMathOperator{\EV}{\mathbb{E}} 

\DeclareMathOperator{\IR}{\mathrm{IR}} 

\DeclareMathOperator{\LR}{\Lambda}

\DeclareMathOperator{\RE}{RE}
\DeclareMathOperator{\ARL}{ARL}

\DeclareMathOperator{\STADD}{STADD}
\DeclareMathOperator{\RIADD}{RIADD}

\DeclareMathOperator*{\arginf}{arg\,inf}

\DeclareMathOperator{\sign}{sign}

\newcommand{\T}{T}

\renewcommand{\le}{\leqslant} 
\renewcommand{\ge}{\geqslant}

\graphicspath{{./gfx/}}

\journal{Communications in Statistics---Simulation and Computation}

\begin{document}

\begin{frontmatter}



\title{\large{\bf\uppercase{On Robustness of the Shiryaev--Roberts Change-Point Detection Procedure under Parameter Misspecification in the Post-Change Distribution}}}


\author{Wenyu\ Du}
\ead{wdu@binghamton.edu}
\ead[url]{http://www.math.binghamton.edu/grads/wdu}
\author{Aleksey\ S.\ Polunchenko\corref{cor-author}}
\ead{aleksey@binghamton.edu}
\ead[url]{http://www.math.binghamton.edu/aleksey}
\author{Grigory\ Sokolov}
\ead{gsokolov@binghamton.edu}
\ead[url]{http://www.math.binghamton.edu/gsokolov}
\cortext[cor-author]{Address correspondence to A.S.\ Polunchenko, Department of Mathematical Sciences, State University of New York (SUNY) at Binghamton, Binghamton, NY 13902--6000, USA; Tel: +1 (607) 777-6906; Fax: +1 (607) 777-2450; Email:~\href{mailto:aleksey@binghamton.edu}{aleksey@binghamton.edu}}
\address{Department of Mathematical Sciences, State University of New York (SUNY) at Binghamton\\Binghamtom, NY 13902--6000, USA}

\begin{abstract}
The gist of the quickest change-point detection problem is to detect the presence of a change in the statistical behavior of a series of sequentially made observations, and do so in an optimal detection-speed-vs.-``false-positive''-risk manner.
When optimality is understood either in the generalized Bayesian sense or as defined in Shiryaev's multi-cyclic setup, the so-called Shiryaev--Roberts (SR) detection procedure is known to be the ``best one can do'', provided, however, that the observations' pre- and post-change distributions are both fully specified.
We consider a more realistic setup, viz.\ one where the post-change distribution is assumed known only up to a parameter, so that the latter may be ``misspecified''. The question of interest is the sensitivity (or robustness) of the otherwise ``best'' SR procedure with respect to a possible misspecification of the post-change distribution parameter.
To answer this question, we provide a case study where, in a specific Gaussian scenario, we allow the SR procedure to be ``out of tune'' in the way of the post-change distribution parameter, and numerically assess the effect of the ``mistuning'' on Shiryaev's (multi-cyclic) Stationary Average Detection Delay delivered by the SR procedure. The comprehensive {\em quantitative} robustness characterization of the SR procedure obtained in the study can be used to develop the respective theory as well as to provide a rational for practical design of the SR procedure. The overall {\em qualitative} conclusion of the study is an expected one: the SR procedure is less (more) robust for less (more) contrast changes and for lower (higher) levels of the false alarm risk.
\end{abstract}

\begin{keyword}


Sequential analysis\sep Shiryaev--Roberts procedure\sep Quality control\sep Quickest change-point detection.
\end{keyword}

\end{frontmatter}

\section{Introduction}
\label{sec:intro}

Sequential (quickest) change-point detection is concerned with the development and evaluation of statistical procedures for rapid and reliable ``on-the-go'' detection of unanticipated changes that may occur in the characteristics of a ``live'' process. Specifically, the process is assumed to be continuously monitored through sequentially made observations (e.g., measurements), and should at any point in time their behavior start to appear as though the process may have (been) statistically changed, the aim is to conclude so within the smallest number of observations possible, subject to a tolerable level of the false positive risk. As soon as such a conclusion is reached, an ``alarm'' is flagged, and an appropriate response action is taken (e.g., an investigation is initiated as to the possible cause of the alarm). For a thorough treatment of the subject's theory as well as for examples of applications, see, e.g.,~\cite{Shiryaev:Book78},~\cite{Basseville+Nikiforov:Book93},~\cite{Poor+Hadjiliadis:Book09}, \cite{Veeravalli+Banerjee:AP2013}, or~\cite[Parts~II~and~III]{Tartakovsky+etal:Book2014}, and the references therein.

A sequential change-point detection procedure is identified with a stopping time, $\T$, that is functionally dependent on the observations, $\{X_n\}_{n\ge1}$; the semantics of $\T$ is that it constitutes a rule whereby one is to stop and declare that the statistical profile of the process under surveillance may have (been) changed. A ``good'' (i.e., optimal or nearly optimal) detection procedure is one that minimizes (or nearly minimizes) the desired detection delay penalty, subject to a constraint on the false alarm risk. For an overview of the major optimality criteria, see, e.g., \cite{Tartakovsky+Moustakides:SA10}, \cite{Polunchenko+Tartakovsky:MCAP2012}, \cite{Veeravalli+Banerjee:AP2013}, \cite{Polunchenko+etal:JSM2013}, or \cite[Part~II]{Tartakovsky+etal:Book2014}.

The basic quickest change-point detection problem assumes\begin{inparaenum}[\itshape(a)]\item\label{lst:assmptn11} that the observations, $\{X_n\}_{n\ge1}$, are independent throughout the entire period of surveillance; \item\label{lst:assmptn22}that the observations' common pre-change probability density function (pdf) $f(x)$ is completely known; and \item that so is the observations' common post-change pdf $g(x)\not\equiv f(x)$\end{inparaenum}. This version of the problem is well-understood and has been solved (either exactly or asymptotically) under a variety of criteria. For a survey of the corresponding state-of-the-art, see, e.g.,~\cite{Tartakovsky+Moustakides:SA10}, \cite{Polunchenko+Tartakovsky:MCAP2012}, \cite{Veeravalli+Banerjee:AP2013}, \cite{Polunchenko+etal:JSM2013}, or~\cite[Part~II]{Tartakovsky+etal:Book2014}.

This work's focus is on a more realistic setup, namely, one where the observations' post-change pdf, $g(x)$, is assumed known only up to a parameter; assumptions {\it(\ref{lst:assmptn11})} and {\it(\ref{lst:assmptn22})} above are retained. While being a more practical assumption, limited (parametric) knowledge of $g(x)$ makes the problem more difficult. In particular, the uncertainty in the post-change distribution parameter opens the door for a possible misspecification thereof. Consequently, when this parameter is set incorrectly (for whatever reason), the performance of any otherwise optimal detection procedure will naturally degrade. The aim of this work is to quantify the severity of this performance degradation as a function of the magnitude of the post-change distribution parameter misspecification and under various levels of the false alarm risk. To make it more clear, our intent in this paper is not to propose a way to deal with the parametric uncertainty in the post-change distribution, but merely to provide a {\em quantitative nonasymptotic} answer to the practically important question of robustness (or sensitivity) of a given detection procedure with respect to possible errors in the value of the post-change distribution parameter.

More concretely, our goal is to examine the robustness question specifically for the so-called Shiryaev--Roberts (SR) procedure, proposed (following quasi-Bayesian considerations) by~\cite{Shiryaev:SMD61,Shiryaev:TPA63} and (independently) by~\cite{Roberts:T66}. The term ``Shiryaev--Roberts'' appears to have been ``coined'' by~\cite{Pollak:AS85}. For a brief account of the SR procedure's history, see, e.g.,~\cite{Pollak:IWSM09}. The interest in the SR procedure is due to two reasons. First, the SR procedure is exactly optimal in Shiryaev's~\citeyearpar{Shiryaev:SMD61,Shiryaev:TPA63} multi-cyclic sense. Second, since Shiryaev's multi-cyclic version of the change-point detection problem is equivalent to the latter's generalized Bayesian setup, the SR procedure is exactly optimal in the generalized Bayesian sense as well. These results were first obtained by~\cite{Shiryaev:SMD61,Shiryaev:TPA63} for a (continuous-time) Brownian motion drift-shift scenario; see also, e.g.,~\cite{Shiryaev:Bachelier2002,Feinberg+Shiryaev:SD2006}. For the discrete-time case, these results were recently established by~\cite{Pollak+Tartakovsky:ISITA2008,Pollak+Tartakovsky:SS09} and by~\cite{Shiryaev+Zryumov:Khabanov2010}. It is noteworthy that neither the celebrated Cumulative Sum (CUSUM) ``inspection scheme'' proposed by~\cite{Page:B54} nor the popular Exponentially Weighted Moving Average (EWMA) chart introduced by~\cite{Roberts:T59} possesses such strong optimality properties. For comparative performance analyses of the SR procedure against the CUSUM scheme and the EWMA chart, see, e.g., \cite{Knoth:FSQC2006}, \cite{Mahmoud+etal:JAS2008}, \cite{Moustakides+etal:CommStat09}, \cite{Tartakovsky+etal:IWSM2009}, and \cite{Polunchenko+etal:ShLnkJAS2013}. However, albeit strong, these optimality properties of the SR procedure hinge on the assumption of complete knowledge of the observations' pre- and post-change distributions, whose pdf's we agreed to denote by $f(x)$ and by $g(x)$, respectively. When the latter involves a parameter that the SR procedure has an incorrect value for, the performance of the SR procedure will no longer be ``best'' (optimal). Furthermore, the larger the error in the parameter value, the bigger the respective performance loss. It is to study this ``cause-and-effect'' relationship  {\em quantitatively} that is the objective of this work, and the focus is entirely on the SR procedure.

Specifically, we offer a case study where, in a particular Gaussian scenario, we intentionally set up the SR procedure so that it ``anticipates'' the observations' pdf to change from $f(x)$ pre-change to $g(x;\tilde{\theta})$ post-change, while the actual change is from $f(x)$ pre-change to $g(x;\theta)$ post-change, where $\tilde{\theta}\neq\theta$, and then assess the effect of the ``mistuning'' on the procedure's performance. The performance characteristics of interest are the usual minimax Average Run Length (ARL) to false alarm proposed by~\cite{Lorden:AMS71} (although see also~\citealt{Page:B54}) and Shiryaev's~\citeyearpar{Shiryaev:SMD61,Shiryaev:TPA63} multi-cyclic Stationary Average Detection Delay (STADD). We compute both metrics numerically, with the aid of an integral-equations-based numerical method that is a hybrid between one proposed and used earlier by~\cite{Moustakides+etal:CommStat09,Moustakides+etal:SS11} and a recent refinement thereof offered and stress-tested by~\cite{Polunchenko+etal:SA2014,Polunchenko+etal:ASMBI2014}. The latter method was designed specifically for the SR procedure, and is more accurate when it comes to evaluation of the ARL to false alarm. The former method is used to compute the STADD, and even though it was originally designed only for the case when $\tilde{\theta}=\theta$, it extends trivially to the case when $\tilde{\theta}\neq\theta$. The study provides an exhaustive quantitative characterization of the SR procedure's robustness. Qualitatively, the overall conclusion of the study is essentially what one would expect it to be: the less (more) contrast the change and the higher (lower) the level of the ARL to false alarm, the less (more) robust the SR procedure devised to detect it. To the best of our knowledge, when the observations are drawn at discrete points in time, no such study has previously been undertaken, even though its significance for applications is apparent. However, for similar studies in the continuous-time case, see, e.g.,~\cite{Pollak+Siegmund:B85} and~\cite{Srivastava+Wu:AS1993}.

The rest of the paper is organized as follows. Firstly, in Section~\ref{sec:problem+background} we formally state the problem, describe the SR procedure, briefly review its properties, and comment on how we intend to evaluate its performance. The case study itself is carried out next, in Section~\ref{sec:case-study}, which is the core section of the paper. Lastly, in Section~\ref{sec:conclusion} we draw conclusions.

\section{The problem and preliminary background}
\label{sec:problem+background}

\begin{figure}[!htb]
    \centering
    \includegraphics[width=0.8\textwidth]{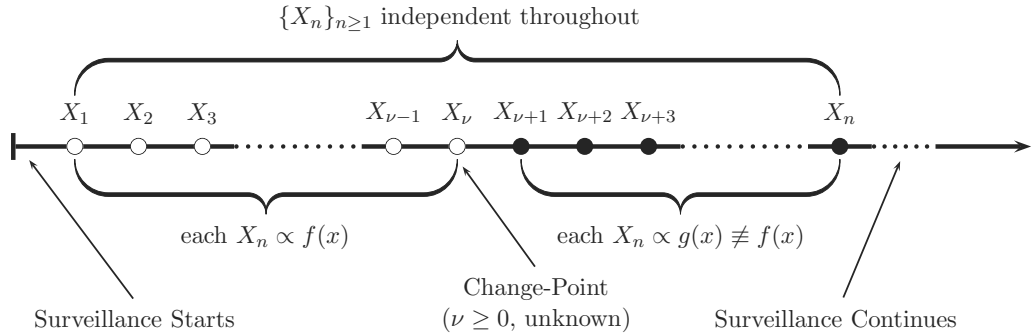}
    \caption{Basic ``minimax-ish'' setup of the quickest change-point detection problem.}
    \label{fig:basic-iid-change-point}
\end{figure}

Since the version of the quickest change-point detection problem that is considered in this work is a build-up to the problem's basic ``minimax'ish'' setup, it is convenient to briefly describe the latter first, i.e., start with the standard assumption of complete knowledge of the observations' pre- and post-change distributions. Let the pdf's of these distributions be $f(x)$ and $g(x)$, respectively; $g(x)\not\equiv f(x)$. Define the change-point, $0\le\nu\le\infty$, as the unknown (but not random) serial index of the final pre-change observation; note that it can potentially be infinite. That is, as illustrated in Figure~\ref{fig:basic-iid-change-point}, the pdf of $X_n$ is $f(x)$ for $1\le n\le\nu$, and $g(x)$ for $n\ge\nu+1$.
The notation $\nu=0$ is to be understood as the case when the pdf of $X_n$ is $g(x)$ for all $n\ge1$, i.e., the distributional pattern of the data, $\{X_n\}_{n\ge1}$, is affected by change {\em ab initio}. Similarly, the notation $\nu=\infty$ is to mean that the pdf of $X_n$ is $f(x)$ for all $n\ge1$, i.e., the distributional pattern of the data, $\{X_n\}_{n\ge1}$, never changes.

Let $\Pr_k$ ($\EV_k$) be the probability measure (expectation) given that the change-point $\nu$ is at time moment (epoch) $k$, i.e., $\nu=k$, where $0\le k\le\infty$. Particularly, $\Pr_\infty$ ($\EV_\infty$) is the probability measure (expectation) when the observations always come from the pdf $f(x)$, i.e., $\nu=\infty$ so that there is never a change. Likewise, $\Pr_0$ ($\EV_0$) is the probability measure (expectation) when the observations' distribution is $g(x)$ from the start (i.e., $\nu=0$).

From now on $\T$ will denote the stopping time associated with a generic detection procedure.

Given this ``minimax-ish'' context, the standard way to gauge the false alarm risk is through Lorden's~\citeyearpar{Lorden:AMS71} ARL to false alarm (see also~\citealt{Page:B54}). The ARL to false alarm is defined as $\ARL(\T)\triangleq\EV_\infty[\T]$ and captures the average number of observations the procedure (given by stopping time $\T$) samples under the pre-change regime before stopping (i.e., before triggering a false alarm). The reciprocal of $\ARL(\T)$ can be interpreted (roughly) as the frequency of false alarms. Hence, the false alarm risk turns out to be inversely proportional to $\ARL(\T)$: the higher (lower) the latter, the lower (higher) the former.

\begin{figure*}[!htp]
    \centering
    \begin{subfigure}{0.9\textwidth}
        \includegraphics[width=\textwidth]{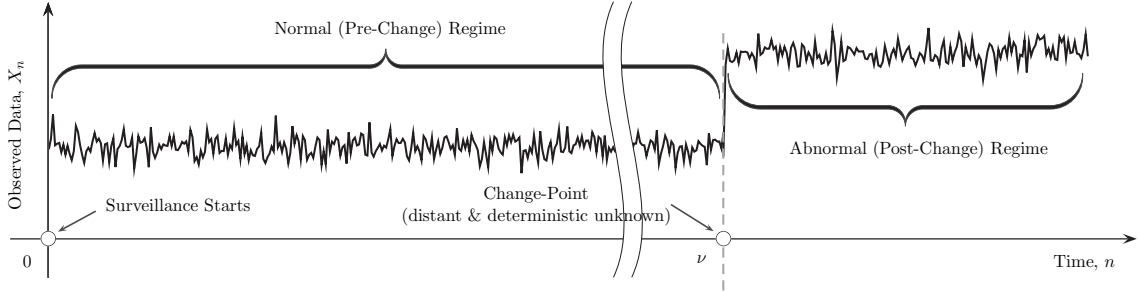}
        \caption{An example of the behavior of a process of interest with a change in mean at time $\nu$.}
    \end{subfigure}
    \begin{subfigure}{0.9\textwidth}
        \includegraphics[width=\textwidth]{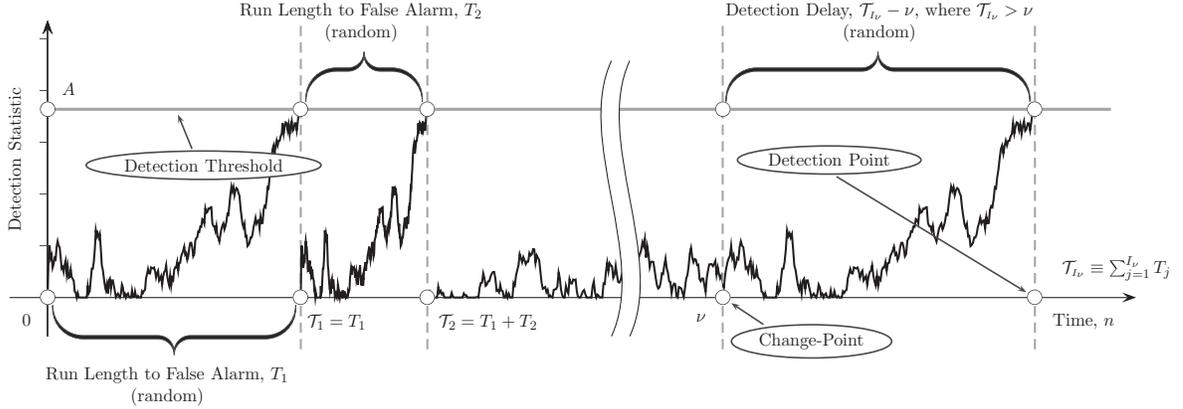}
        \caption{Typical behavior of the detection statistic in the multi-cyclic mode.}
    \end{subfigure}
    \caption{Multi-cyclic change-point detection in a stationary regime.}
    \label{fig:multi-cyclic-idea}
\end{figure*}

To introduce the multi-cyclic change-point detection problem, let
\begin{align*}
\Delta(\gamma)
&\triangleq
\Bigl\{\T\colon\ARL(\T)\ge\gamma\Bigr\},\;\text{where}\;\gamma>1,
\end{align*}
denote the class of procedures (stopping times), $\T$, whose the ARL to false alarm is at least $\gamma>1$, a pre-selected tolerance level. Suppose now that it is of utmost importance to detect the change as quickly as possible, even at the expense of raising many false alarms (using a repeated application of the same procedure) before the change occurs. Put otherwise, in exchange for the assurance that the change will be detected with maximal speed, one agrees to go through a ``flurry'' of false alarms along the way (the false alarms are ensued from repeatedly applying the same procedure, starting from scratch after each false alarm). This scenario is shown in Figure~\ref{fig:multi-cyclic-idea}.

Formally, let $T_1,T_2,\ldots$ be sequential independent repetitions of the same stopping time, $\T$, and let ${\cal T}_j\triangleq T_1+T_2+\cdots+T_j$, $j\ge1$, be the time of the $j$-th alarm. Define $I_\nu\triangleq\min\{j\ge1\colon {\cal T}_j>\nu\}$ so that ${\cal T}_{\scriptscriptstyle I_\nu}$ is the time of detection of a true change that occurs at time moment $\nu$ after $I_\nu-1$ false alarms had been raised. One can then view the difference ${\cal T}_{\scriptscriptstyle I_\nu}-\nu(\ge0)$ as the detection delay. Let
\begin{align*}
\STADD(\T)
&\triangleq
\lim_{\nu\to\infty}\EV_\nu[{\cal T}_{\scriptscriptstyle I_\nu}-\nu]
\end{align*}
be the limiting value of the Average Detection Delay (ADD) referred to as the {\em Stationary ADD} (STADD); this metric was introduced by~\cite{Shiryaev:SMD61,Shiryaev:TPA63}. The multi-cyclic change-point detection problem then is:
\begin{align}\label{eq:multi-cyclic-problem}
\text{to find}\;
\T_{\mathrm{opt}}
&=\arginf_{\T\in\Delta(\gamma)}\STADD(\T)\;\text{for any given}\;\gamma>1;
\end{align}
cf.~\cite{Shiryaev:SMD61,Shiryaev:TPA63}.

We note that, in this setup, $\ARL(\T)$ is exactly the average distance between successive false alarms. Therefore, $1/\ARL(\T)$ can be thought as the frequency of false alarms. Since higher (lower) frequency of false alarms clearly corresponds to higher (lower) false alarm risk, lower (higher) levels of $\ARL(\T)$ correspond to higher (lower) levels of the false alarm risk.

As can be gathered from the description, the ``intrinsic assumption'' of the multi-cyclic change-point detection problem is that the process under surveillance is not expected to be affected by change ``for a while'', i.e., the change-point, $\nu$, is large. Scenarios where this is a reasonable assumption may be encountered, e.g., in cybersecurity (see, e.g.,~\citealt{Polunchenko+etal:SA2012} or \citealt{Tartakovsy+etal:IEEE-JSTSP2013}) and in the economic design of quality control charts (see, e.g.,~\citealt{Duncan:JASA1956,Montgomery:JQT1980,Lorenzen+Vance:T1986,Ho+Case:JQT1994}).

The multi-cyclic change-point detection problem~\eqref{eq:multi-cyclic-problem} has been solved by~\cite{Pollak+Tartakovsky:ISITA2008,Pollak+Tartakovsky:SS09} and by~\cite{Shiryaev+Zryumov:Khabanov2010} who showed that the solution is the Shiryaev--Roberts (SR) procedure, due to~\cite{Shiryaev:SMD61,Shiryaev:TPA63} and~\cite{Roberts:T66}. We reiterate that neither the CUSUM scheme nor the EWMA chart possesses such a strong optimality property. As a matter of fact, the comparative performance analyses carried out by~\cite{Moustakides+etal:CommStat09,Tartakovsky+etal:IWSM2009,Polunchenko+etal:ShLnkJAS2013} confirmed experimentally that both the CUSUM scheme and the EWMA chart are inferior to (i.e., are outperformed by) the SR procedure in the multi-cyclic sense. For similar analyses carried out in continuous time, see, e.g., \cite{Pollak+Siegmund:B85} and \cite{Srivastava+Wu:AS1993}.

\begin{figure}[!htp]
    \centering
    \includegraphics[width=0.8\textwidth]{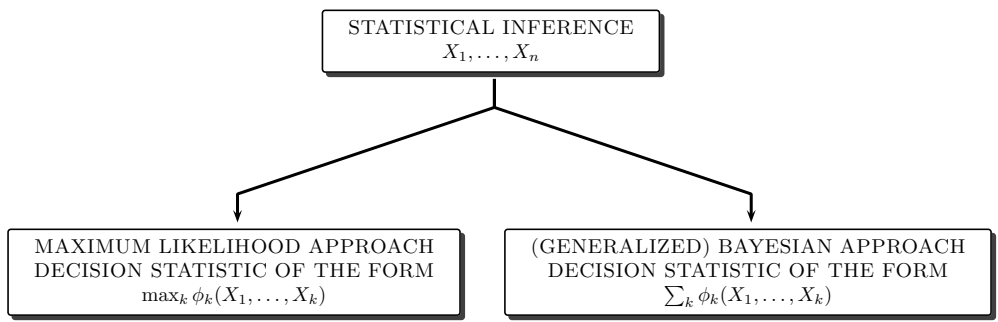}
    \caption{Two principally different approaches to statistical inference.}
    \label{fig:stat-inference-approaches}
\end{figure}

To introduce the SR procedure, let us first point out that, by and large, in all of statistics, there are two principally different ways to extract information from data to make inference. One of these approaches is based on maximization of a certain functional of the data. The other approach is based on aggregation of information in a sum-like manner. This is illustrated in Figure~\ref{fig:stat-inference-approaches}.

In change-point detection, the first approach is manifested in the CUSUM scheme, which uses the maximum-likelihood principle to ``sense'' whether the process under surveillance has gone ``out-of-control''. The EWMA chart is an example of a detection procedure whose decision statistic is based on aggregation of information in a Bayesian-like manner. The SR procedure is also of the latter type. It is formally defined as the stopping time
\begin{align}\label{eq:T-SR-def}
\mathcal{S}_A
&\triangleq
\inf\big\{n\ge1\colon R_n\ge A\big\},\;\text{such that}\;\inf\{\varnothing\}=\infty,
\end{align}
where $A>0$ is a detection threshold (control limit) used to control the level of the ARL to false alarm, and
\begin{align}\label{eq:Rn-SR-def}
R_n
&\triangleq
\sum_{k=1}^n \prod_{i=k}^n\LR_i,\; n\ge1,
\end{align}
is the SR detection statistic; here and onward $\LR_i\triangleq g(X_i)/f(X_i)$ denotes the ``instantaneous'' likelihood ratio (LR) for the $i$-th data point $X_i$. Note the recursion
\begin{align}\label{SRstatrec}
R_{n+1}
&=
(1+R_{n})\LR_{n+1}\;\text{for}\; n=0,1,\ldots\;\text{with}\; R_0=0.
\end{align}

An important property of the SR statistic $\{R_n\}_{n\ge0}$ is that the sequence $\{R_n-n\}_{n\ge0}$ is a zero-mean $\Pr_\infty$-martingale, i.e., $\EV_\infty[R_n-n]=0$ for any $n\ge0$. From this and Doob's optional stopping (sampling) theorem (see, e.g.,~\citealt[Chapter~VII]{Shiryaev:Book1995},~\citealt[Subsection~2.3.2]{Poor+Hadjiliadis:Book09} or~\citealt[Theorem~2.3.1,~p.~31]{Tartakovsky+etal:Book2014}), one can conclude that $\EV_\infty[R_{\mathcal{S}_A}-\mathcal{S}_A]=0$ so that $\ARL(\mathcal{S}_A)\triangleq\EV_\infty[\mathcal{S}_A]=\EV_\infty[R_{\mathcal{S}_A}]\ge A$. Hence, for any given $\gamma>1$, if $A\ge\gamma$, then $\ARL(\mathcal{S}_A)\ge\gamma$ and $\mathcal{S}_A\in\Delta(\gamma)$. As argued by~\cite{Kenett+Pollak:JAP1996}, such a simple relation between $\ARL(\mathcal{S}_A)$ and $A$ endows the SR procedure with certain data-analytic advantages over the CUSUM scheme and the EWMA chart. Furthermore, note that all this is valid irrespective of the particular $f(x)$ and $g(x)$.

A more accurate connection between $\ARL(\mathcal{S}_A)$ and $A$ has been established by~\cite{Pollak:AS87} who showed that $\ARL(\mathcal{S}_A)=(A/\zeta)[1+o(1)]$ as $A\to\infty$, i.e., $\ARL(\mathcal{S}_A)\approx A/\zeta$ for sufficiently large $A$. To define $\zeta$, let $S_n\triangleq\sum_{i=1}^n\log\LR_i$, $n\ge1$, and let $\tau_a\triangleq\inf\{n\ge1\colon S_n\ge a\}$, $a>0$ (again, with the understanding that $\inf\{\varnothing\}=\infty$). Then $\kappa_a\triangleq S_{\tau_a}-a\,(\ge0)$ is the so-called ``overshoot'' (excess over the level $a>0$ at stopping), and $\zeta\triangleq\lim_{a\to\infty}\EV_0[e^{-\kappa_a}]$, and is referred to as the ``limiting average exponential overshoot''. In general, $\zeta$ is between $0$ and $1$, and the evaluation of this model-dependent constant falls within the scope of nonlinear renewal theory; see, e.g.,~\cite{Woodroofe:Book82},~\cite[Section~II.C]{Veeravalli+Banerjee:AP2013} or~\cite[Section~2.6]{Tartakovsky+etal:Book2014}.

More importantly, as was mentioned earlier, according to the result of~\cite{Pollak+Tartakovsky:ISITA2008,Pollak+Tartakovsky:SS09} and also to that of~\cite{Shiryaev+Zryumov:Khabanov2010}, the SR procedure is {\em exactly} $\STADD(\T)$-optimal. That is, formally: $\STADD(\mathcal{S}_{A_\gamma})=\inf_{\T\in\Delta(\gamma)}\STADD(\T)$ for every $\gamma>1$, where $A_\gamma>0$ is the solution of the equation $\ARL(\mathcal{S}_{A_\gamma})=\gamma>1$. However, and we mentioned that already as well, this result ceases to hold when the SR statistic is ``out of tune'' in the way of either $f(x)$ or $g(x)$. We are interested in the case when $g(x)$ is given parametrically as $g(x;\theta)$, and the value of $\theta$ that is used by the SR statistic is $\tilde{\theta}\neq\theta$. To reflect that in this case $\STADD(\mathcal{S}_{A_\gamma})$ is a function of both $\tilde{\theta}$ and $\theta$, we will use the notation $\STADD_{\tilde{\theta},\theta}(\mathcal{S}_{A_\gamma})$. Then clearly $\STADD_{\tilde{\theta},\theta}(\mathcal{S}_{A_\gamma})\ge\STADD_{\theta,\theta}(\mathcal{S}_{A_\gamma})$ for all $\tilde{\theta}\neq\theta$. The aim of this work is to assess the sensitivity of $\STADD_{\tilde{\theta},\theta}(\mathcal{S}_{A_\gamma})$ with respect to the magnitude of the difference between $\tilde{\theta}$ and $\theta$ and under various levels of the ARL to false alarm $\ARL(\mathcal{S}_{A_\gamma})=\gamma>1$.

It is also noteworthy that the SR procedure~\eqref{eq:T-SR-def}--\eqref{SRstatrec} is optimal in the generalized Bayesian sense as well. Specifically, the generalized Bayesian change-point detection problem is: to find $\T_{\mathrm{opt}}=\arginf_{\T\in\Delta(\gamma)}\RIADD(\T)$ for any given $\gamma>1$, where $\RIADD(\T)$ is the {\em Relative Integral ADD} defined as
\begin{align}\label{eq:RIADD-def}
\RIADD(\T)
&\triangleq
\dfrac{1}{\ARL(\T)}\sum_{\nu=0}^{\infty}\EV_{\nu}[\max\{0,\T-\nu\}].
\end{align}

This version of the quickest change-point detection problem is a limiting case of the classical Bayesian change-point detection problem considered and solved by~\cite{Shiryaev:SMD61,Shiryaev:TPA63}. Specifically, the limit is with respect to the prior distribution of the change-point, $\nu$, which is assumed to be improper uniform, so that  each time instance becomes equally likely to be the change-point.

The exact generalized Bayesian optimality the SR procedure can be formally stated as follows: $\RIADD(\mathcal{S}_{A_\gamma})=\inf_{\T\in\Delta(\gamma)}\RIADD(\T)$ for every $\gamma>1$, where $A_\gamma>0$ is again the solution of the equation $\ARL(\mathcal{S}_{A_\gamma})=\gamma>1$. This result is also due to~\cite{Pollak+Tartakovsky:ISITA2008,Pollak+Tartakovsky:SS09} and~\cite{Shiryaev+Zryumov:Khabanov2010}, and the proof is based on first showing that the generalized Bayesian version of the quickest change-point detection problem and its multi-cyclic setup~\eqref{eq:multi-cyclic-problem} are equivalent in the sense that $\RIADD(\T)\equiv\STADD(\T)$ for any stopping time $\T$, and then exploiting the fact that the SR procedure is exactly $\STADD(\T)$-optimal.

The fact that $\RIADD(\T)\equiv\STADD(\T)$ for any stopping time $\T$ was used by~\cite{Moustakides+etal:SS11} to develop a numerical method to compute $\STADD(\mathcal{S}_A)$, i.e., the STADD delivered by the SR procedure~\eqref{eq:T-SR-def}--\eqref{SRstatrec}. Specifically, they observed that when $\T$ is based off of a Markovian detection statistic (note that the SR statistic $\{R_n\}$ {\em is} Markovian), the infinite sum appearing in the right-hand side of~\eqref{eq:RIADD-def} is a convergent geometric series. This allowed~\cite{Moustakides+etal:SS11} to derive an integral (renewal) equation directly on the entire sum without any truncation, and then develop a numerical method to treat the integral equation; their method was recently extended by~\cite{Polunchenko+etal:ASMBI2014}. This is precisely how we intend to evaluate $\STADD(\mathcal{S}_A)$ in our case study offered in the next section: $\STADD(\mathcal{S}_A)$ will be computed numerically as $\RIADD(\mathcal{S}_A)$ with the aid of the numerical methods of~\cite{Moustakides+etal:SS11} and~\cite{Polunchenko+etal:ASMBI2014}. We note that these numerical methods do not require any truncation of either the infinite sum appearing in the definition of $\RIADD(\T)$ or the limit appearing in the definition of $\STADD(\T)$. This fact makes both methods more accurate.

We would like to conclude the introduction of the SR procedure with a remark pertaining to a possible interpretation of the SR statistic, $\{R_n\}_{n\ge0}$. Part of the reason why the SR procedure---in spite of its simplicity and strong optimality properties---has not yet received proper attention from engineers and applied scientists (in particular, from the quality control community), is that the SR statistic is problematic to chart. Specifically, the problem is that the decision-making mechanism behind the SR procedure is not as clearly understood as that behind, e.g., the CUSUM scheme or the EWMA chart, which have {\em de facto} been the industry ``workhorse'' for decades now. A simple intuitive explanation of how the SR statistic ``works'', i.e., a meaningful ``engineering'' interpretation of the SR statistic, could help bridge this gap, and, in the long run, might also help ``pave the way'' for the SR procedure into the ``engineering world''. An attempt to provide one such interpretation was previously made by~\cite{Kenett+Pollak:JAP1996} who, in view of the quasi-Bayesian nature of the SR procedure, argued that the SR statistic, $\{R_n\}_{n\ge0}$, may be regarded as a $p$-value.

The interpretation we have in mind for the SR statistic lends the latter a financial flavor, and involves an ordinary savings account and basic financial interest theory. Suppose first that one opens a savings account at a bank and immediately makes a deposit of $\$\,1$, so that at ``time zero'' (also referred to as ``now'' or ``today'') the balance is $\$\,1$.
\begin{wrapfigure}{R}{0.55\textwidth}
    \centering
    \includegraphics[width=0.5\textwidth]{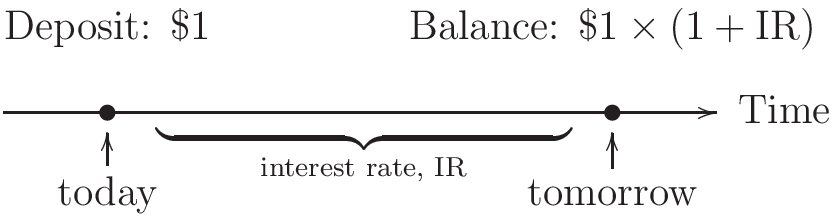}
    \caption{Financial analogy to interpret the SR statistic.}
    \label{fig:SR_R_deposit_one_period}       
\end{wrapfigure}
If the account accrues interest at a rate of $\mathrm{IR}$ (decimals) per time period, then it is easy to see that the balance in the account ``tomorrow'' (i.e., in one time period from ``today'') will be $\$\,1\times(1+\mathrm{IR})$. This is schematically illustrated in Figure~\ref{fig:SR_R_deposit_one_period}.

\begin{figure}[!htp]
    \centering
    \includegraphics[width=0.95\textwidth]{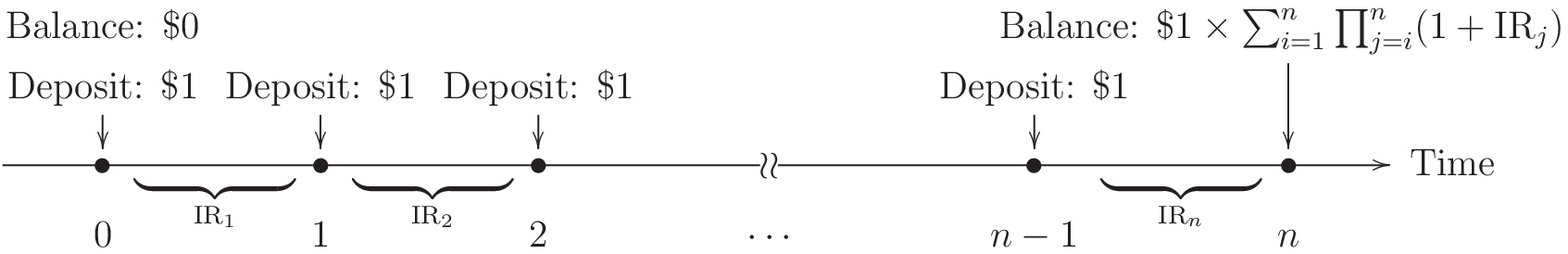}
    \caption{Financial analogy to interpret the SR statistic.}
    \label{fig:SR_R_deposit_multiple_periods}       
\end{figure}

Now suppose that the time horizon is longer than just ``one day'', and once the bank account is opened, one funds it not with one, but with a series of {\em periodic} single-dollar deposits. Moreover, suppose also that the interest rate {\em varies} from period to period. This is shown in Figure~\ref{fig:SR_R_deposit_multiple_periods}.

Then, at time $n$ before the $n$-th deposit is made, the balance will be
\begin{align*}
B_n
&=
\$\,1\times\sum_{i=1}^{n}
\prod_{j=i}^n (1+\mathrm{IR}_j),\;n\ge1,
\end{align*}
and we hasten to note the similarity between this formula and that for the SR statistic~\eqref{eq:Rn-SR-def}: the two formulae are identical if the LR at the $n$-th epoch, $\LR_n$, is regarded as the $n$-th period interest factor, i.e., the quantity $(1+\IR_n)$. This suggests that the SR statistic, $\{R_n\}_{n\ge0}$, can be thought of the balance in one's savings account at epoch $n\ge1$, assuming that the account is\begin{inparaenum}[\itshape(a)]\item funded through a series of periodic one-dollar deposits, and \item credited (compound) interest every period\end{inparaenum}. If the detection threshold, $A>0$, is interpreted as a ``target balance'', then the SR stopping time $\mathcal{S}_A$ is the amount of time one is to wait before the account accumulates the target balance. Clearly, the higher the interest rate, the shorter the wait. This can now be used to explain how the SR statistic makes its decision as to the presence of a change in the observed process.

On the one hand, since $\EV_\infty[\LR_n]=1$ for all $n\ge1$, this is equivalent to saying that $\IR_n$ is (on average) zero, and, therefore, if there is no change, then there is no interest, and the value of money does not change over time, so the balance in the account after $n$ single-dollar deposits is simply the combined amount thereof, which is $\$\,n$. This is in perfect alignment with the well-known $\Pr_\infty$-martingale property of the SR procedure mentioned above, i.e., that $\EV_\infty[R_n-n]=0$ for any $n\ge0$.

On the other hand, since $\EV_0[\LR_n]>1$ for all $n\ge1$, this is the same as to say that $\IR_n$ is (on average) positive, and, therefore, the account is credited (compound) interest, so that the balance grows faster. Hence, if there is a change, the target balance of $\$A>0$ is achieved sooner, resulting in a quicker detection of the change.

We would like to stress that this interpretation of the SR statistic is not to say that the SR procedure is ``better'' or ``worse'' than, e.g., the CUSUM scheme or the EWMA chart. The point is merely to help the ``uninitiated'' to intuitively understand why it is not unreasonable to at least expect the SR approach to work as a change-point detection tool to begin with.
With this in mind, it is also instructional to mention that the decision-making process behind the CUSUM scheme is drastically different from that behind the SR procedure. Specifically, recall first that the CUSUM chart calls for stopping at $\mathcal{C}_A\triangleq\inf\{n\ge1\colon W_n\ge A\}$, where $A>0$ is again the detection threshold (control limit) and $\{W_n\}_{n\ge0}$ is the CUSUM statistic defined as
\begin{align*}
W_n
&\triangleq
\max\{0,W_{n-1}+\log\LR_n\}\;\text{for}\; n=0,1,\ldots\;\text{with}\; W_0=0;
\end{align*}
cf.~\cite{Page:B54}. Then, on the one hand, observe that because $\EV_\infty[\log\LR_n]<0$ for all $n\ge1$, it follows that under the ``no-change'' regime the CUSUM statistic has a negative drift, which causes the CUSUM statistic to constantly ``gravitate'' toward zero. Since the latter is a reflective barrier for the CUSUM statistic, this endows the CUSUM scheme with a built-in mechanism to reset its statistic when, after a while of surveillance, the data have given no reason to believe their distribution changed. This internal resetting mechanism plays a major role in CUSUM's minimax optimality first established by~\cite{Moustakides:AS86} and then also proved by~\cite{Ritov:AS90} using a different approach. By contrast, the SR procedure does not operate in this manner, and the SR statistic has no resetting feature to it. On the other hand, observe that $\EV_{\nu}[\log\LR_n]>0$ for all $n\ge1$ and $0\le \nu<n$, and therefore, as soon as the observations are affected by change, the drift of the CUSUM statistic becomes positive, so that the CUSUM statistic starts to climb up toward the detection threshold, and eventually either hits it or surpasses it, signalling an alarm to indicate the change has (apparently) occurred. This difference in the way the CUSUM statistic behaves in the pre- and in the post-change regimes makes the CUSUM statistic very convenient to interpret when it is plotted against time: not only will the plot show that the change is likely to have occurred, but it will also provide a clue as to when it occurred. The SR statistic does not offer this kind of convenience.

\begin{figure}[!htp]
    \centering
    \begin{subfigure}{0.45\textwidth}
        \includegraphics[width=\textwidth]{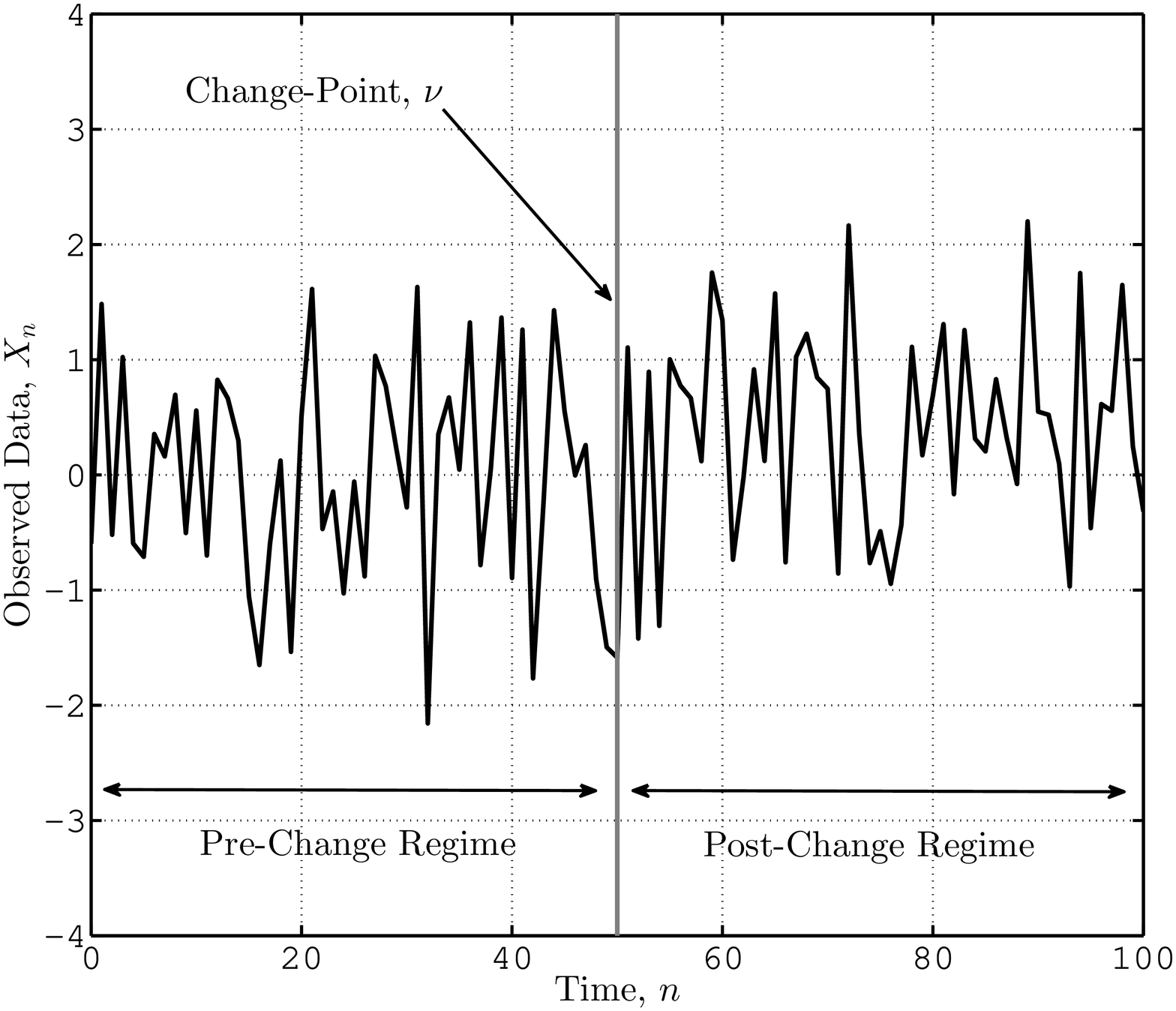}
        \caption{Observed data, $\{X_n\}$.}
        \label{fig:cs-vs-sr-data}
    \end{subfigure}\\
    \begin{subfigure}{0.4\textwidth}
        \includegraphics[width=\textwidth]{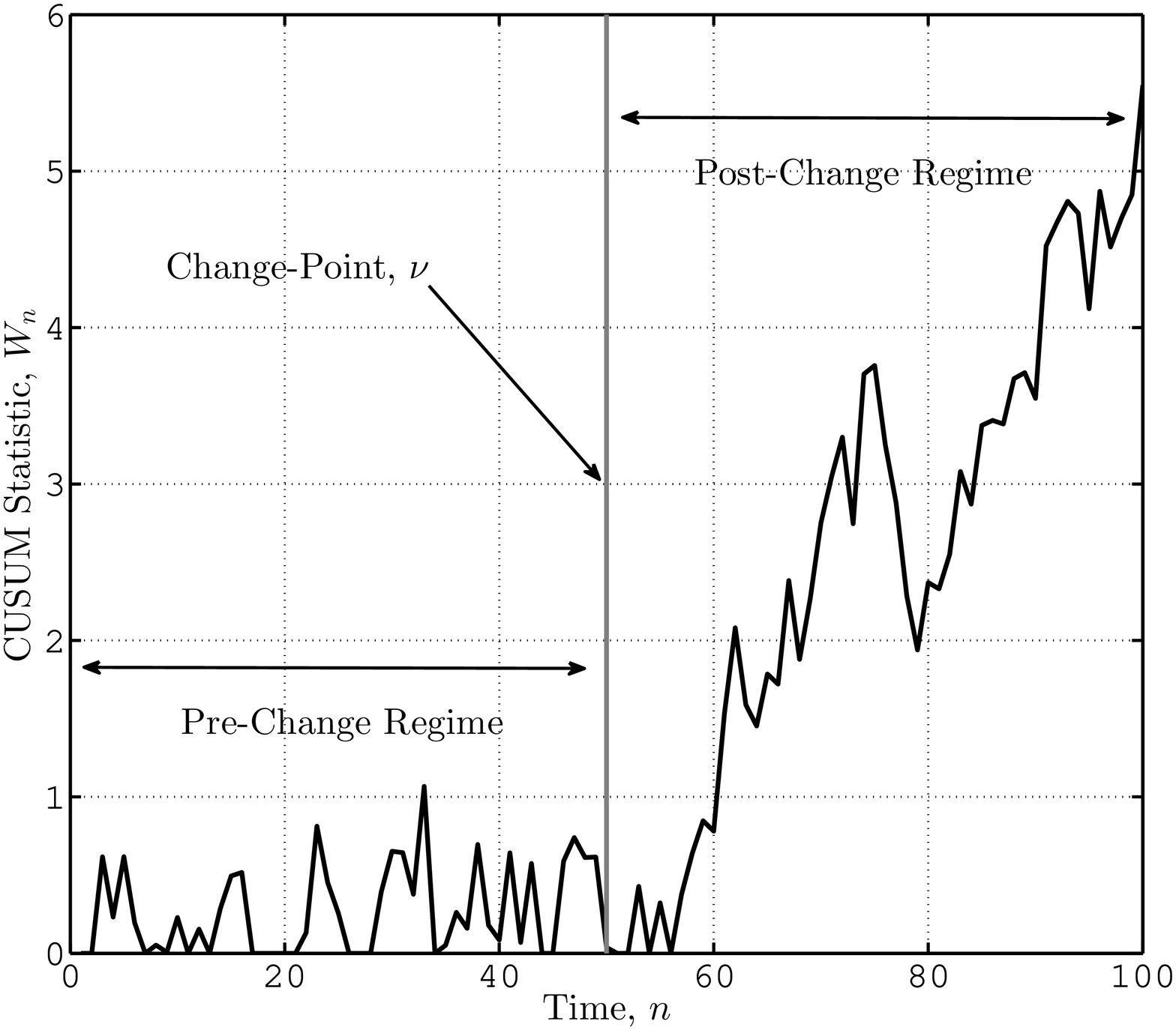}
        \caption{CUSUM statistic, $\{W_n\}$.}
        \label{fig:cs-vs-sr-cusum}
    \end{subfigure}
    \;\;
    \begin{subfigure}{0.4\textwidth}
        \includegraphics[width=\textwidth]{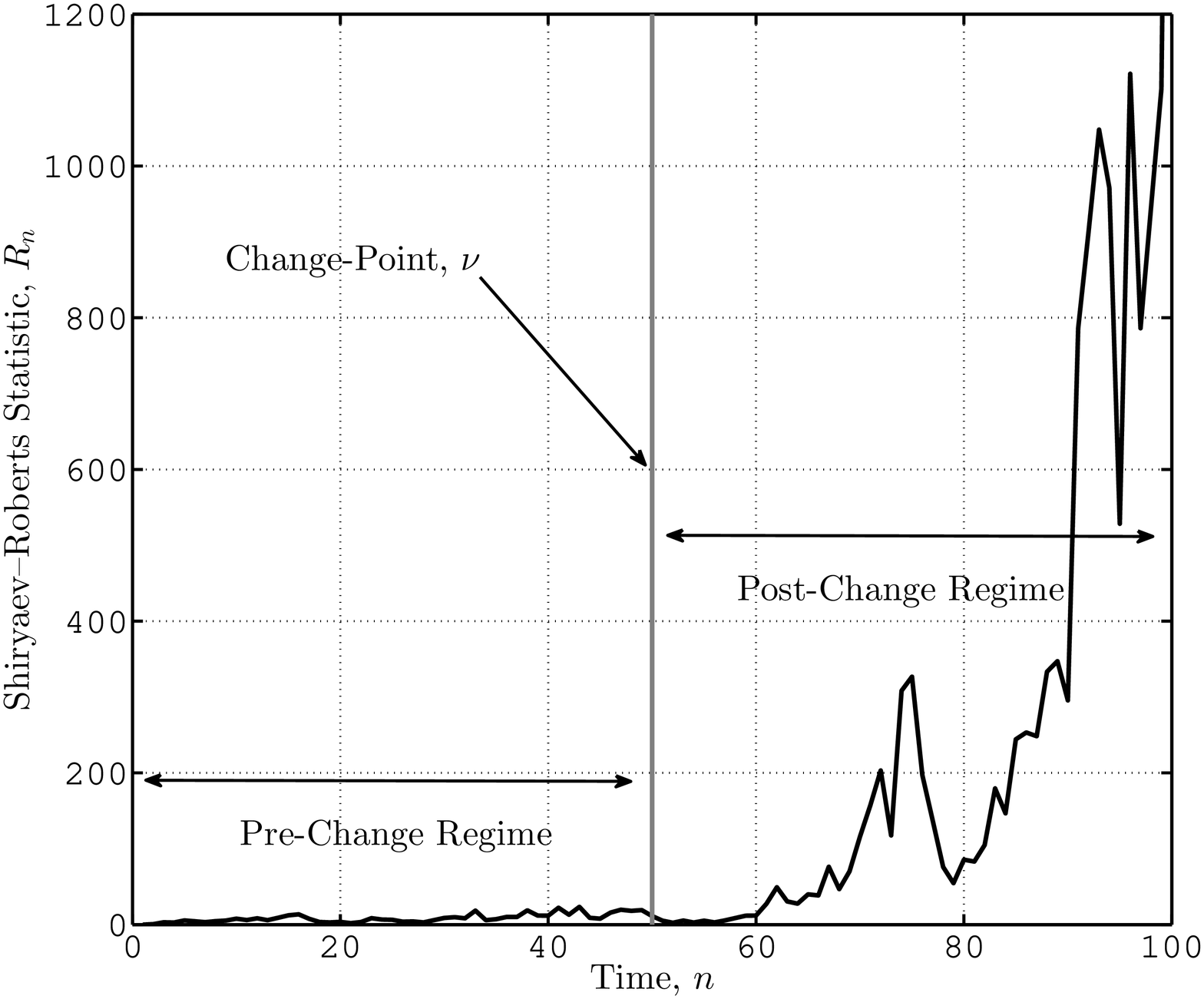}
        \caption{SR statistic, $\{R_n\}$.}
        \label{fig:cs-vs-sr-sr}
    \end{subfigure}
    \caption{Illustration of the behavior of the CUSUM statistic and that of the SR statistic for the same data.}
    \label{fig:cs-vs-sr}
\end{figure}

To illustrate the above point, consider Figure~\ref{fig:cs-vs-sr} which gives an example of the typical behavior of the CUSUM statistic $\{W_n\}$ and that of the SR statistic $\{R_n\}$ for the problem of detecting a shift in the mean of a series of independent standard Gaussian observations $\{X_n\}$. Specifically, Figure~\ref{fig:cs-vs-sr-data} presents a sample trajectory of $\{X_n\}$ of $100$ points, with the change occurring at epoch $\nu=50$, i.e., right down the middle of the sample, and the magnitude of the shift is 0.5. Shown underneath Figure~\ref{fig:cs-vs-sr-data} are Figures~\ref{fig:cs-vs-sr-cusum} and~\ref{fig:cs-vs-sr-sr} which depict the respective trajectories of the CUSUM statistic $\{W_n\}$ and of the SR statistic $\{R_n\}$.

As can be seen, both the CUSUM statistic and the SR statistic change their behavior drastically as soon as the data series undergoes a change in the mean. Both statistics begin to climb up, and if there were a detection threshold (control limit) imposed, one could declare the change as having occurred as soon as the respective statistic would reach or exceed the threshold. However, as was explained above, at the local level the two statistics are poles apart: they each ``sense'' the presence of the change in a completely different manner.

\section{The case study}
\label{sec:case-study}

As the centerpiece of this work, this section is devoted to a case study, where, in a concrete scenario and subject to the constraint $\ARL(\mathcal{S}_A)=\gamma$ for a given $\gamma>1$, we quantify $\STADD_{\tilde{\theta},\theta}(\mathcal{S}_A)$ as a function of $\tilde{\theta}$ and $\theta$, and, in particular, assess the sensitivity of $\STADD_{\tilde{\theta},\theta}(\mathcal{S}_A)$ with respect to the magnitude of the difference $\tilde{\theta}-\theta$. We would like to reiterate that to compute $\ARL(\mathcal{S}_A)$ and $\STADD_{\tilde{\theta},\theta}(\mathcal{S}_A)$, we will rely on two numerical methods: one proposed by~\cite{Moustakides+etal:SS11} and its refinement due to~\cite{Polunchenko+etal:SA2014,Polunchenko+etal:ASMBI2014}.

Specifically, suppose that the observations, $\{X_n\}_{n\ge1}$, are independent and Gaussian-distributed, with mean zero pre-change and with mean $\theta\neq0$ post-change, i.e., $\theta$ is the actual (true) value of the post-change mean; suppose also that the variance is 1 and does not change. Formally, the pre- and post-change pdf's in this case are
\begin{align*}
f(x)
&=
\dfrac{1}{\sqrt{2\pi}}\exp\left\{-\dfrac{x^2}{2}
\right\}\;\text{and}\;
g(x;\theta)=\dfrac{1}{\sqrt{2\pi}}\exp\left\{-\dfrac{(x-\theta)^2}{2}\right\},
\end{align*}
respectively, where $x\in\mathbb{R}$ and $\theta\neq0$. When it is assumed that $\theta$ has no way to be misspecified, this Gaussian scenario is the standard ``testbed'' model widely used in the literature. We are interested in the case when $\theta$ {\em can} be misspecified (for whatever reason), and let $\tilde{\theta}$ denote the respective putative (or ``anticipated'') value of $\theta$. That is, it is $\tilde{\theta}$ (and not $\theta$) that is to be used to construct the corresponding LR to then base the SR statistic off of.

For the Gaussian scenario under consideration, the corresponding ``instantaneous'' LR for the $n$-th data point, $X_n$, i.e., $\LR_n\triangleq g(X_n;\tilde{\theta})/f(X_n)$, can be seen to be
\begin{align*}
\LR_n
&=
\exp\left\{\tilde{\theta} X_n-\frac{\tilde{\theta}^2}{2}\right\},\;n\ge1,
\end{align*}
and, therefore, for each $n\ge1$, the LR's distribution is log-normal and such that $\log\LR_n$ has mean $-\tilde{\theta}^2/2$ and variance $\tilde{\theta}^2$ under measure $\Pr_\infty$, and mean $\tilde{\theta}\theta-\tilde{\theta}^2/2\,\,(\neq-\tilde{\theta}^2/2)$ and variance $\tilde{\theta}^2$ under measure $\Pr_0(\cdot|\theta)$, where $\Pr_0(\cdot|\theta)$ denotes the probability measure under the assumption that the change is in effect from the start and that the actual distribution of the observations is $g(x;\theta)$. Specifically, let $P_{\infty}^{\LR}(t)\triangleq\Pr_\infty(\LR_1\le t)$, $t\ge0$, and $P_0^{\LR}(t|\tilde{\theta},\theta)\triangleq\Pr_0(\LR_1\le t|\theta)$, $t\ge0$, be the cumulative distribution functions (cdf's) of the LR in the pre- and in the post-change regimes, respectively. Then, we obtain:
\begin{empheq}[%
    left={%
        P_{\infty}^{\LR}(t)=
    \empheqlbrace}]{align*}
&\Phi\left(\sign(\tilde{\theta})\left[\frac{1}{\tilde{\theta}}\log t+\frac{\tilde{\theta}}{2}\right]\right),\;\;\text{for $t\ge0$;}\\
&0,\;\;\text{for $t<0$,}
\end{empheq}
and
\begin{empheq}[%
    left={%
        P_{0}^{\LR}(t|\tilde{\theta},\theta)=%
    \empheqlbrace}]{align*}
&\Phi\left(\sign(\tilde{\theta})\left[\frac{1}{\tilde{\theta}}\log t+\frac{\tilde{\theta}}{2}-\theta\right]\right),\;\;\text{for $t\ge0$;}\\
&0,\;\;\text{for $t<0$,}
\end{empheq}
where
\begin{align*}
\Phi(x)
&\triangleq
\frac{1}{\sqrt{2\pi}}\int_{-\infty}^x e^{-\tfrac{t^2}{2}}\,dt,\; x\in\mathbb{R},
\end{align*}
is the standard Gaussian cdf. We note that, as expected, $P_{\infty}^{\LR}(t)$ does not depend on $\theta$. With $P_\infty^{\LR}(t)$ and $P_0^{\LR}(t|\tilde{\theta},\theta)$ both expressed explicitly, the numerical method of~\cite{Moustakides+etal:SS11} and that of~\cite{Polunchenko+etal:SA2014,Polunchenko+etal:ASMBI2014} can be implemented. We implemented the methods in MATLAB, the popular software package for scientific simulation and computation developed by MathWorks, Inc. Both methods were set up so that the accuracy is on the order of a fraction of a percent.

We are now in a position to begin our case study. To that end, the obvious point of departure is the question: How does one choose the detection threshold, $A>0$, so as to guarantee that $\ARL(\mathcal{S}_A)=\gamma$ for a given $\gamma>1$? To answer this question, recall that, as was discussed in Section~\ref{sec:problem+background}, at least when $A>0$ is sufficiently large, $\ARL(\mathcal{S}_A)\approx A/\zeta$, where $\zeta$ is the limiting average exponential overshoot. Thus, if $\zeta$ were known (at least approximately), then $A\approx\zeta\gamma$ would provide (approximately) the sought value of the detection threshold needed to ensure $\ARL(\mathcal{S}_A)\approx\gamma$ for a given $\gamma>1$. For our Gaussian scenario, $\zeta=\zeta(\tilde{\theta})$ can be computed numerically from the well-known (exact) formula:
\begin{align*}
\zeta
&=
\dfrac{2}{\tilde{\theta}^2}\exp\left\{-2\sum_{k=1}^\infty\frac{1}{k}\Phi\left(-\dfrac{\tilde{\theta}}{2}\sqrt{k}\right)\right\},
\end{align*}
where $\Phi(x)$ is the standard Gaussian cdf; see, e.g.,~\cite[Example~3.1,~pp.~32--33]{Woodroofe:Book82} or~\cite[Section~3.1.5,~p.~137]{Tartakovsky+etal:Book2014}. Since it is well known that
\begin{align*}
2\Phi(-t)
&\ge
\sqrt{\frac{2}{\pi}}\dfrac{t}{1+t^2} e^{-\tfrac{t^2}{2}},\;\text{for all}\;t>0,
\end{align*}
it is easy to see that the infinite series appearing under the exponent in the above formula for $\zeta$ converges exponentially fast. Hence, to compute $\zeta$ accurately, it is ``safe'' to simply ``chop off'' the series at a reasonably distant term. We used the first $10^{6}$ terms, which is substantially more than needed to compute $\zeta$ very accurately. We computed and tabulated $\zeta$ for $\tilde{\theta}=0.1,0.2,\ldots,1.0$. The obtained values (rounded to the first six decimal places) are reported in the second leftmost column in Table~\ref{tab:Threshold_vs_theta_ARL}. These values coincide with those previously found by~\cite[p.~33]{Woodroofe:Book82}. With $\zeta$ known, the detection threshold, $A=A_{\gamma}$, needed to ensure $\ARL(\mathcal{S}_{A_{\gamma}})\approx\gamma$ for a given $\gamma>1$ can be determined as $A_{\gamma}=\zeta\gamma$. However, note that the latter formula is an approximate one (it becomes exact only asymptotically, when $A\to\infty$), and if $A_{\gamma}=\zeta\gamma$ and is a {\em finite} number, then the actual proximity of $\ARL(\mathcal{S}_{A_{\gamma}})$ to $\gamma$ needs to be verified. To that end, for greater certainty, we used the numerical method of~\cite{Polunchenko+etal:SA2014} and computed numerically (with high accuracy) the actual level of the ARL to false alarm for each value of the detection threshold. These numerically computed ARL to false alarm levels can be considered trustworthy. Table~\ref{tab:Threshold_vs_theta_ARL} reports the obtained results. Specifically, each column under the heading ``Desired level of the Average Run Length (ARL) to false alarm'' corresponds to a specific level of the ARL to false alarm $\gamma>1$ picked between $10^2$ and $10^4$. Each row corresponds to a specific value of $\tilde{\theta}$ (the value of $\theta$ in this case is irrelevant). The content of each cell is two numbers placed one on top of the other. The top number is the detection threshold computed as $A=\gamma\zeta$ for the corresponding $\gamma$ and $\tilde{\theta}$. The bottom number (in parentheses) is the actual level of the ARL to false alarm (computed numerically using the numerical method of~\citealt{Polunchenko+etal:SA2014}) that corresponds to that particular detection threshold. The main conclusion that can be made from the results is that the approximation $\ARL(\mathcal{S}_A)\approx A/\zeta$ is quite accurate, uniformly across all $\tilde{\theta}$ and all $\gamma>1$.
\renewcommand{\arraystretch}{1.1}
\begin{sidewaystable}
    \centering\renewcommand\tabcolsep{3pt}
    \caption{Characterization of the detection threshold ($A>0$) for selected values of the post-change mean ($\tilde{\theta}=\theta$) and ARL to false alarm (i.e., \ $\ARL(\mathcal{S}_A)=\gamma>1$).}
    \scalebox{0.7}{
    \begin{tabularx}{1.2\textheight}{@{}l@{\quad}l|c|*{11}{Y}@{}}
    \toprule
      & & &\multicolumn{11}{c}{Desired level of Average Run Length (ARL) to false alarm (i.e., $\ARL(\mathcal{S}_A)=\gamma>1$)}\\[1mm]
      & & $\zeta$ &$1\times10^2$ & $2\times10^2$ & $3\times10^2$ & $4\times10^2$ & $5\times10^2$ & $6\times10^2$ & $7\times10^2$ & $8\times10^2$ & $9\times10^2$
        & $1\times10^3$ & $1\times10^4$\\
    \cmidrule[0.6pt]{1-14}
            \multirow{22}{*}{\rotatebox[origin=c]{90}{Post-change mean (putative$=$true)}}
            &\multirow{2}{*}{0.1}
            &\multirow{2}{*}{0.943408}
            & $94.34$ & $188.68$ & $283.02$ & $377.36$ & $471.7$ & $566.04$ & $660.38$ & $754.72$ & $849.06$  & $943.4$ & $9,434.08$\\
            & &
            & $(100.28)$ & $(200.28)$ & $(300.28)$ & $(400.28)$ & $(500.28)$ & $(600.28)$ & $(700.28)$ & $(800.27)$ & $(900.27)$ & $(1,000.27)$ & $(10,000.28)$\\
    \cmidrule{2-14}
            &\multirow{2}{*}{0.2}
            &\multirow{2}{*}{0.890037}
            & $89.0$ & $178.00$ & $267.01$ & $356.01$ & $445.01$ & $534.02$ & $623.02$ & $712.02$ & $801.03$ & $890.03$ & $8,900.36$\\
            & &
            & $(100.31)$ & $(200.31)$ & $(300.32)$ & $(400.31)$ & $(500.31)$ & $(600.31)$ & $(700.31)$ & $(800.31)$ & $(900.31)$ & $(1,000.31)$ & $(10,000.3)$\\
    \cmidrule{2-14}
            &\multirow{2}{*}{0.3}
            &\multirow{2}{*}{0.839721}
            & $83.97$ & $167.94$ & $251.91$ & $335.88$ & $419.86$ & $503.83$ & $587.8$ & $671.77$ & $755.74$ & $839.72$ & $8,397.21$ \\
            & &
            & $(100.35)$ & $(200.35)$ & $(300.35)$ & $(400.35)$ & $(500.35)$ & $(600.35)$ & $(700.35)$ & $(800.35)$ & $(900.34)$ & $(1,000.35)$ & $(10,000.36)$\\
    \cmidrule{2-14}
            &\multirow{2}{*}{0.4}
            &\multirow{2}{*}{0.792298}
            & $79.22$ & $158.45$ & $237.68$ & $316.91$ & $396.14$ & $475.37$ & $554.6$ & $633.83$ & $713.06$ & $792.29$ & $7,922.98$\\
            & &
            & $(100.39)$ & $(200.39)$ & $(300.39)$ & $(400.39)$ & $(500.39)$ & $(600.39)$ & $(700.39)$ & $(800.39)$ & $(900.39)$ & $(1,000.39)$ & $(10,000.4)$\\
    \cmidrule{2-14}
            &\multirow{2}{*}{0.5}
            &\multirow{2}{*}{0.747615}
            & $74.76$ & $149.52$ & $224.28$ & $299.04$ & $373.8$ & $448.56$ & $523.33$ & $598.09$ & $672.85$ & $747.61$ & $7,476.15$\\
            & &
            & $(100.45)$ & $(200.44)$ & $(300.44)$ & $(400.44)$ & $(500.44)$ & $(600.44)$ & $(700.45)$ & $(800.44)$ & $(900.44)$ & $(1,000.44)$ & $(10,000.45)$\\
    \cmidrule{2-14}
            &\multirow{2}{*}{0.6}
            &\multirow{2}{*}{0.705525}
            & $70.55$ & $141.1$ & $211.65$ & $282.2$ & $352.76$ & $423.31$ & $493.86$ & $564.41$ & $634.97$ & $705.52$ & $7,055.25$ \\
            & &
            & $(100.5)$ & $(200.5)$ & $(300.5)$ & $(400.5)$ & $(500.5)$ & $(600.5)$ & $(700.5)$ & $(800.5)$ & $(900.5)$ & $(1000.5)$ & $(10,000.5)$\\
    \cmidrule{2-14}
            &\multirow{2}{*}{0.7}
            &\multirow{2}{*}{0.665887}
            & $66.58$ & $133.17$ & $199.76$ & $266.35$ & $332.94$ & $399.53$ & $466.12$ & $532.7$ & $599.29$ & $665.88$ & $6,658.87$\\
            & &
            & $(100.55)$ & $(200.55)$ & $(300.55)$ & $(400.56)$ & $(500.57)$ & $(600.56)$ & $(700.56)$ & $(800.55)$ & $(900.55)$ & $(1,000.55)$ & $(10,000.56)$\\
    \cmidrule{2-14}
            &\multirow{2}{*}{0.8}
            &\multirow{2}{*}{0.628566}
            & $62.85$ & $125.71$ & $188.56$ & $251.42$ & $314.28$ & $377.13$ & $439.99$ & $502.85$ & $565.7$ & $628.56$ & $6,285.66$\\
            & &
            & $(100.62)$ & $(200.62)$ & $(300.61)$ & $(400.62)$ & $(500.62)$ & $(600.61)$ & $(700.62)$ & $(800.62)$ & $(900.61)$ & $(1,000.62)$ & $(10,000.62)$ \\
    \cmidrule{2-14}
            &\multirow{2}{*}{0.9}
            &\multirow{2}{*}{0.593435}
            & $59.34$ & $118.68$ & $178.03$ & $237.37$ & $296.71$ & $356.06$ & $415.4$ & $474.74$ & $534.09$ & $593.43$ & $5,934.35$\\
            & &
            & $(100.7)$ & $(200.7)$ & $(300.7)$ & $(400.7)$ & $(500.7)$ & $(600.7)$ & $(700.7)$ & $(800.69)$ & $(900.7)$ & $(1,000.7)$ & $(10,000.71)$\\
    \cmidrule{2-14}
            &\multirow{2}{*}{1.0}
            &\multirow{2}{*}{0.56037}
            & $56.03$ & $112.07$ & $168.11$ & $224.14$ & $280.18$ & $336.22$ & $392.25$ & $448.29$ & $504.33$ & $560.37$ & $5,603.7$ \\
            & &
            & $(100.77)$ & $(200.78)$ & $(300.79)$ & $(400.77)$ & $(500.78)$ & $(600.78)$ & $(700.77)$ & $(800.78)$ & $(900.78)$ & $(1,000.79)$ & $(10,000.78)$\\
    \bottomrule
    \end{tabularx}
    } 
    \label{tab:Threshold_vs_theta_ARL}
\end{sidewaystable}

We are now in a position to examine the sensitivity of $\STADD_{\tilde{\theta},\theta}(\mathcal{S}_A)$ with respect to $\theta-\tilde{\theta}$, i.e., the size of the error in the post-change mean. Specifically, in view of the fact that $\STADD_{\tilde{\theta},\theta}(\mathcal{S}_A)$ is the lowest when $\theta-\tilde{\theta}=0$, it makes sense to use $\STADD_{\theta,\theta}(\mathcal{S}_A)$ as a benchmark and measure the performance loss via the following relative efficiency (RE) metric:
\begin{align*}
\RE_{\tilde{\theta},\theta}(\mathcal{S}_A)
&\triangleq
\dfrac{\STADD_{\tilde{\theta},\theta}(\mathcal{S}_A)-\STADD_{{\theta},{\theta}}(\mathcal{S}_A)}{\STADD_{{\theta},{\theta}}(\mathcal{S}_A)},
\end{align*}
where it is understood that the constraint $\ARL(\mathcal{S}_A)=\gamma$ is fulfilled for a given $\gamma>1$. Due to the exact STADD-optimality of the SR procedure when $\tilde{\theta}=\theta$, it is clear that $\RE_{\tilde{\theta},\theta}(\mathcal{S}_A)\ge0$ in general, and that $\RE_{\theta,\theta}(\mathcal{S}_A)=0$ in particular. The actual robustness of the SR procedure would then be inversely proportional to $\RE_{\tilde{\theta},\theta}(\mathcal{S}_A)$: smaller (higher) values of $\RE_{\tilde{\theta},\theta}(\mathcal{S}_A)$ would correspond to greater (lower) robustness. We will consider three levels of the ARL to false alarm: $\gamma=10^2$, which corresponds to high false alarm risk, $\gamma=10^3$, which corresponds to moderate false alarm risk, and $\gamma=10^4$, which corresponds to low false alarm risk.

Let us first consider the case when $\ARL(\mathcal{S}_A)=\gamma=10^2$. The values of the detection threshold for different values of $\tilde{\theta}$ are given in Table~\ref{tab:Threshold_vs_theta_ARL}. Table~\ref{tab:STADD_r0_vs_theta__gamma1e2} summarizes of the behavior of $\STADD_{\tilde{\theta},\theta}(\mathcal{S}_A)$ and that of $\RE_{\tilde{\theta},\theta}(\mathcal{S}_A)$ each regarded as a bivariate function of $\tilde{\theta}$ and $\theta$. Specifically, the content of each cell in the table is a pair of numbers, one on top of the other: the top number is the actual $\STADD_{\tilde{\theta},\theta}(\mathcal{S}_A)$-value computed for the respective pair of $\tilde{\theta}$ and $\theta$ and rounded to two decimal places, and the bottom number (given in parentheses) is the respective $\RE_{\tilde{\theta},\theta}(\mathcal{S}_A)$-value given as a percentage. For example, the top number in cell positioned at the intersection of the first (topmost) row and the rightmost column is $9.86$. This number is the value of $\STADD_{\tilde{\theta},\theta}(\mathcal{S}_A)$ when $\tilde{\theta}=0.1$ and $\theta=1.0$ (and the ARL to false alarm is $10^2$). The number underneath it is $80.68\%$, and it is the relative excess of the $9.86$ over the ideal-case performance. The ideal case is when $\tilde{\theta}=\theta=1.0$, and the respective value of $\STADD_{\tilde{\theta},\theta}(\mathcal{S}_A)$ is $5.46$ as can be inferred from the content of the cell positioned at the intersection of the bottommost row and the rightmost column. The $80.68\,\%$ is computed as $100\,\%\times(9.86-5.46)/5.46$. This explains why all the percentages along the table's diagonal are zero: it is the ideal case, i.e., when $\tilde{\theta}=\theta$, and in this case the SR procedure is exactly $\STADD(\T)$-optimal.
\begin{table}[!htb]
    \centering
    \caption{Characterization of $\STADD_{\tilde{\theta},\theta}(\mathcal{S}_A)$ and $\RE_{\tilde{\theta},\theta}(\mathcal{S}_A)$ as functions of the putative value ($\tilde{\theta}$) and of the true value ($\theta$) of the post-change mean for the ARL to false alarm of $10^2$ (i.e., $\ARL(\mathcal{S}_A^r)=\gamma=10^2$).}
    \scalebox{0.65}{
    \begin{tabularx}{1.5\textwidth}{@{}l@{\quad}l|*{10}{Y}@{}}
    \toprule
      & & \multicolumn{10}{c}{True value of the post-change mean ($\theta$)}\\[2mm]
      & & 0.1 & 0.2 & 0.3 & 0.4 & 0.5 & 0.6 & 0.7 & 0.8 & 0.9 & 1.0\\
    \cmidrule[0.6pt]{1-12}%
            \multirow{24}{*}{\rotatebox[origin=c]{90}{Putative value of the post-change mean ($\tilde{\theta}$)}}
            &\multirow{2}{*}{0.1}
            &40.14 & 29.7 & 23.58 & 19.57 & 16.75 & 14.67 & 13.05 & 11.77 & 10.73 & 9.86\\
            &
            &$(0\,\%)$ & $(5.18\,\%)$ & $(14.29\,\%)$ & $(24.2\,\%)$ & $(34.18\,\%)$ & $(44.0\,\%)$ & $(53.58\,\%)$ & $(62.89\,\%)$ & $(71.92\,\%)$ & $(80.68\,\%)$\\
\cmidrule{2-12}
            &\multirow{2}{*}{0.2}
            &41.93 & 28.24 & 21.01 & 16.67 & 13.81 & 11.79 & 10.3 & 9.14 & 8.23 & 7.49\\
            &
            &$(4.47\,\%)$ & $(0\,\%)$ & $(1.87\,\%)$ & $(5.8\,\%)$ & $(10.59\,\%)$ & $(15.78\,\%)$ & $(21.12\,\%)$ & $(26.52\,\%)$ & $(31.89\,\%)$ & $(37.21\,\%)$\\
\cmidrule{2-12}
            &\multirow{2}{*}{0.3}
            &44.58 & 28.74 & 20.63 & 15.92 & 12.92 & 10.86 & 9.36 & 8.24 & 7.36 & 6.66\\
            &
            & $(11.06\,\%)$ & $(1.78\,\%)$ & $(0\,\%)$ & $(1.03\,\%)$ & $(3.4\,\%)$ & $(6.6\,\%)$ & $(10.17\,\%)$ & $(13.99\,\%)$ & $(17.93\,\%)$ & $(21.93\,\%)$\\
\cmidrule{2-12}
            &\multirow{2}{*}{0.4}
            &47.1 & 29.74 & 20.84 & 15.76 & 12.57 & 10.43 & 8.91 & 7.77 & 6.9 & 6.21\\
            &
            & $(17.35\,\%)$ & $(5.31\,\%)$ & $(1.05\,\%)$ & $(0\,\%)$ & $(0.69\,\%)$ & $(2.42\,\%)$ & $(4.77\,\%)$ & $(7.52\,\%)$ & $(10.52\,\%)$ & $(13.67\,\%)$\\
\cmidrule{2-12}
            &\multirow{2}{*}{0.5}
            &49.41 & 30.93 & 21.34 & 15.88 & 12.49 & 10.24 & 8.66 & 7.5 & 6.61 & 5.92\\
            &
            & $(23.09\,\%)$ & $(9.52\,\%)$ & $(3.47\,\%)$ & $(0.74\,\%)$ & $(0\,\%)$ & $(0.52\,\%)$ & $(1.85\,\%)$ & $(3.73\,\%)$ & $(5.96\,\%)$ & $(8.43\,\%)$\\
\cmidrule{2-12}
            &\multirow{2}{*}{0.6}
            &51.51 & 32.2 & 22.01 & 16.17 & 12.56 & 10.18 & 8.53 & 7.34 & 6.43 & 5.73\\
            &
            & $(28.32\%)$ & $(14.05\,\%)$ & $(6.7\,\%)$ & $(2.58\,\%)$ & $(0.58\,\%)$ & $(0\,\%)$ & $(0.41\,\%)$ & $(1.49\,\%)$ & $(3.03\,\%)$ & $(4.9\,\%)$\\
\cmidrule{2-12}
            &\multirow{2}{*}{0.7}
            &53.43 & 33.52 & 22.79 & 16.58 & 12.74 & 10.23 & 8.5 & 7.25 & 6.32 & 5.6\\
            &
            & $(33.12\,\%)$ & $(18.72\,\%)$ & $(10.46\,\%)$ & $(5.2\,\%)$ & $(2.06\,\%)$ & $(0.47\,\%)$ & $(0\,\%)$ & $(0.33\,\%)$ & $(1.23\,\%)$ & $(2.54\,\%)$\\
\cmidrule{2-12}
            &\multirow{2}{*}{0.8}
            &55.21 & 34.85 & 23.64 & 17.08 & 13.02 & 10.36 & 8.53 & 7.23 & 6.26 & 5.52\\
            &
            & $(37.55\,\%)$ & $(23.43\,\%)$ & $(14.6\,\%)$ & $(8.39\,\%)$ & $(4.24\,\%)$ & $(1.7\,\%)$ & $(0.39\,\%)$ & $(0\,\%)$ & $(0.28\,\%)$ & $(1.04\,\%)$\\
\cmidrule{2-12}
            &\multirow{2}{*}{0.9}
            &56.86 & 36.18 & 24.55 & 17.66 & 13.36 & 10.55 & 8.62 & 7.25 & 6.24 & 5.47\\
            &
            & $(41.65\,\%)$ & $(28.13\,\%)$ & $(19.01\,\%)$ & $(12.04\,\%)$ & $(6.98\,\%)$ & $(3.56\,\%)$ & $(1.44\,\%)$ & $(0.34\,\%)$ & $(0\,\%)$ & $(0.23\,\%)$\\
\cmidrule{2-12}
            &\multirow{2}{*}{1.0}
            &58.39 & 37.5 & 25.49 & 18.29 & 13.76 & 10.79 & 8.76 & 7.32 & 6.26 & 5.46\\
            &
            & $(45.48\,\%)$ & $(32.79\,\%)$ & $(23.59\,\%)$ & $(16.05\,\%)$ & $(10.19\,\%)$ & $(5.94\,\%)$ & $(3.05\,\%)$ & $(1.25\,\%)$ & $(0.3\,\%)$ & $(0\,\%)$\\
    \bottomrule
    \end{tabularx}
    } 
    \label{tab:STADD_r0_vs_theta__gamma1e2}
\end{table}

\begin{figure}[!htp]
    \centering
    \includegraphics[width=0.95\textwidth]{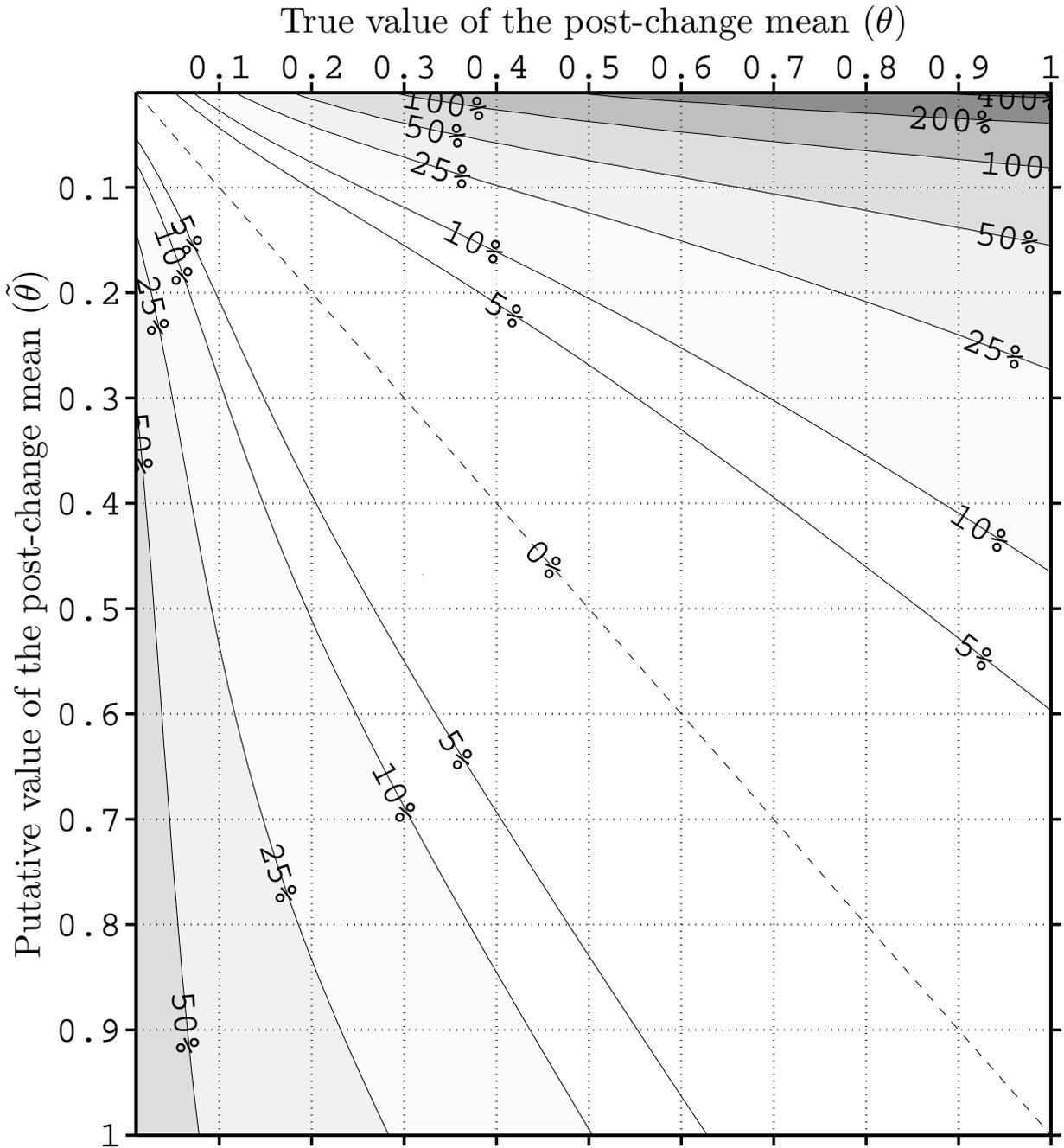}
    \caption{Characterization of $\RE_{\tilde{\theta},\theta}(\mathcal{S}_A)$ as a function of the putative value ($\tilde{\theta}$) and of the true value ($\theta$) of the post-change mean for the Gaussian scenario for the ARL to false alarm of $10^2$ (i.e., $\ARL(\mathcal{S}_A^r)=\gamma=10^2$).}
    \label{fig:STADD_r0_vs_theta__gamma1e2}
\end{figure}

As a complement to Table~\ref{tab:STADD_r0_vs_theta__gamma1e2}, Figure~\ref{fig:STADD_r0_vs_theta__gamma1e2} provides a graphical representation of $\RE_{\tilde{\theta},\theta}(\mathcal{S}_A)$ for various values of $\tilde{\theta}$ and $\theta$ by means of a grayscale ``heat diagram''. The idea is to represent lower (higher) values of $\RE_{\tilde{\theta},\theta}(\mathcal{S}_A)$ with fainter (stronger) shades of gray. For better perception, Figure~\ref{fig:STADD_r0_vs_theta__gamma1e2} also contains contours corresponding to a few selected levels of $\RE_{\tilde{\theta},\theta}(\mathcal{S}_A)$. For instance, the curves labeled ``5\%'' provide boundaries for the values of $\tilde{\theta}$ and $\theta$ for which $\RE_{\tilde{\theta},\theta}(\mathcal{S}_A)\le 5\%$. One may note that more or less uniformly across the entire Table~\ref{tab:STADD_r0_vs_theta__gamma1e2} (i.e., uniformly in $\tilde{\theta}$ and $\theta$), if the error in the post-change mean is about $0.1\div0.2$ in absolute value, then the respective relative performance degradation is on the order of a few percentage points, which would be considered ``tolerable'' in most practical situations. Figure~\ref{fig:STADD_r0_vs_theta__gamma1e2} confirms this. One may also note that the sensitivity is higher when $\theta$ and $\tilde{\theta}$ are small, and then it gradually decreases as $\theta$ and $\tilde{\theta}$ increase. This is especially apparent in Figure~\ref{fig:STADD_r0_vs_theta__gamma1e2}, where the ``5\,\%''-level boundaries are narrower around the origin, and get further apart as one moves away from the origin. We also note that when $|\theta-\tilde{\theta}|\ge0.5$, the performance loss may be hard to ignore in practice, as it is on the order of tens of percent.

\begin{figure}[!htp]
    \centering
    \includegraphics[width=0.95\textwidth,keepaspectratio=true]{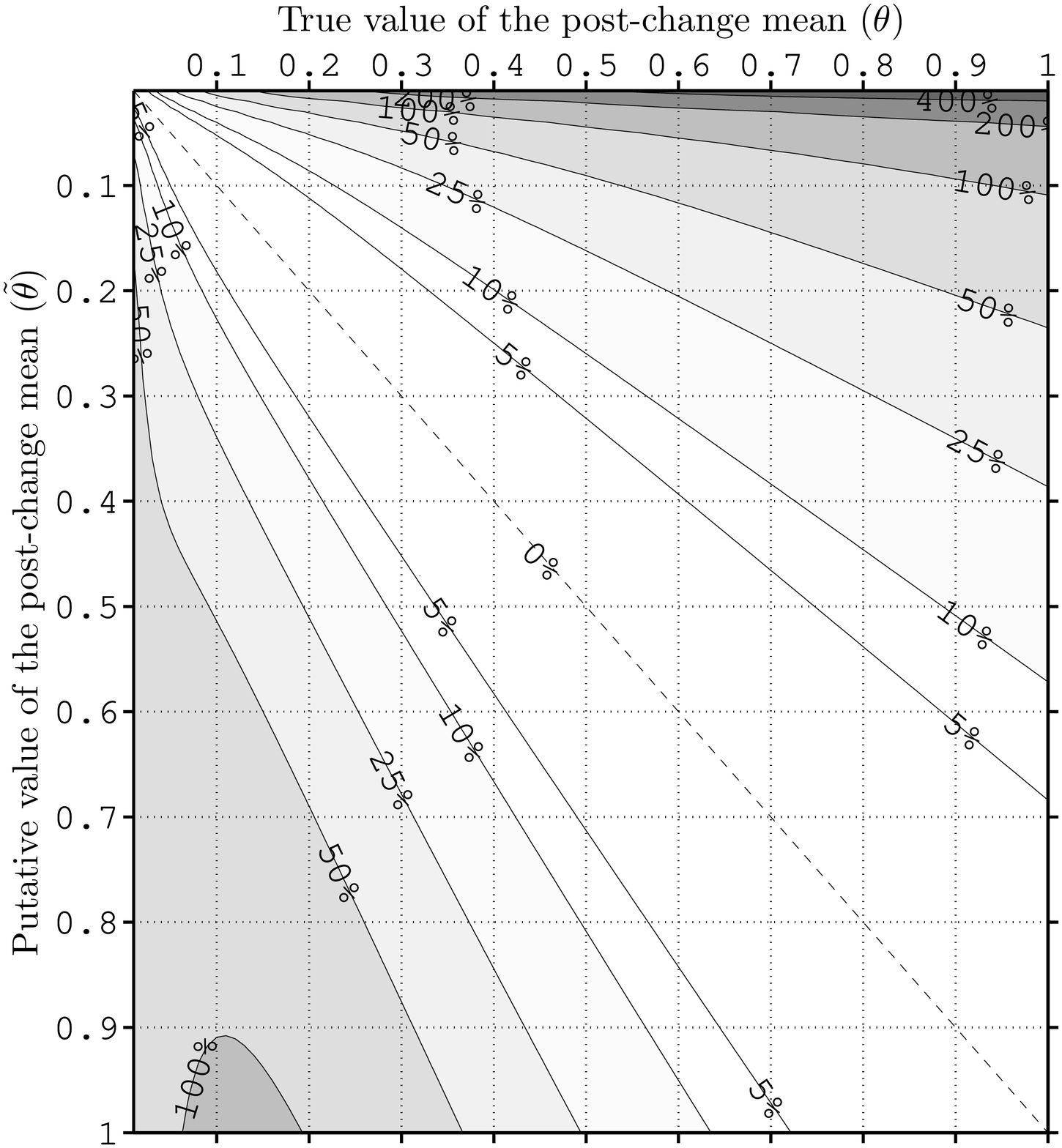}
    \caption{Characterization of $\RE_{\tilde{\theta},\theta}(\mathcal{S}_A)$ as a function of the putative value ($\tilde{\theta}$) and of the true value ($\theta$) of the post-change mean for the ARL to false alarm of $10^3$ (i.e., $\ARL(\mathcal{S}_A^r)=\gamma=10^3$).}
    \label{fig:STADD_r0_vs_theta__gamma1e3}
\end{figure}

\begin{figure}[!htp]
    \centering
    \includegraphics[width=0.95\textwidth,keepaspectratio=true]{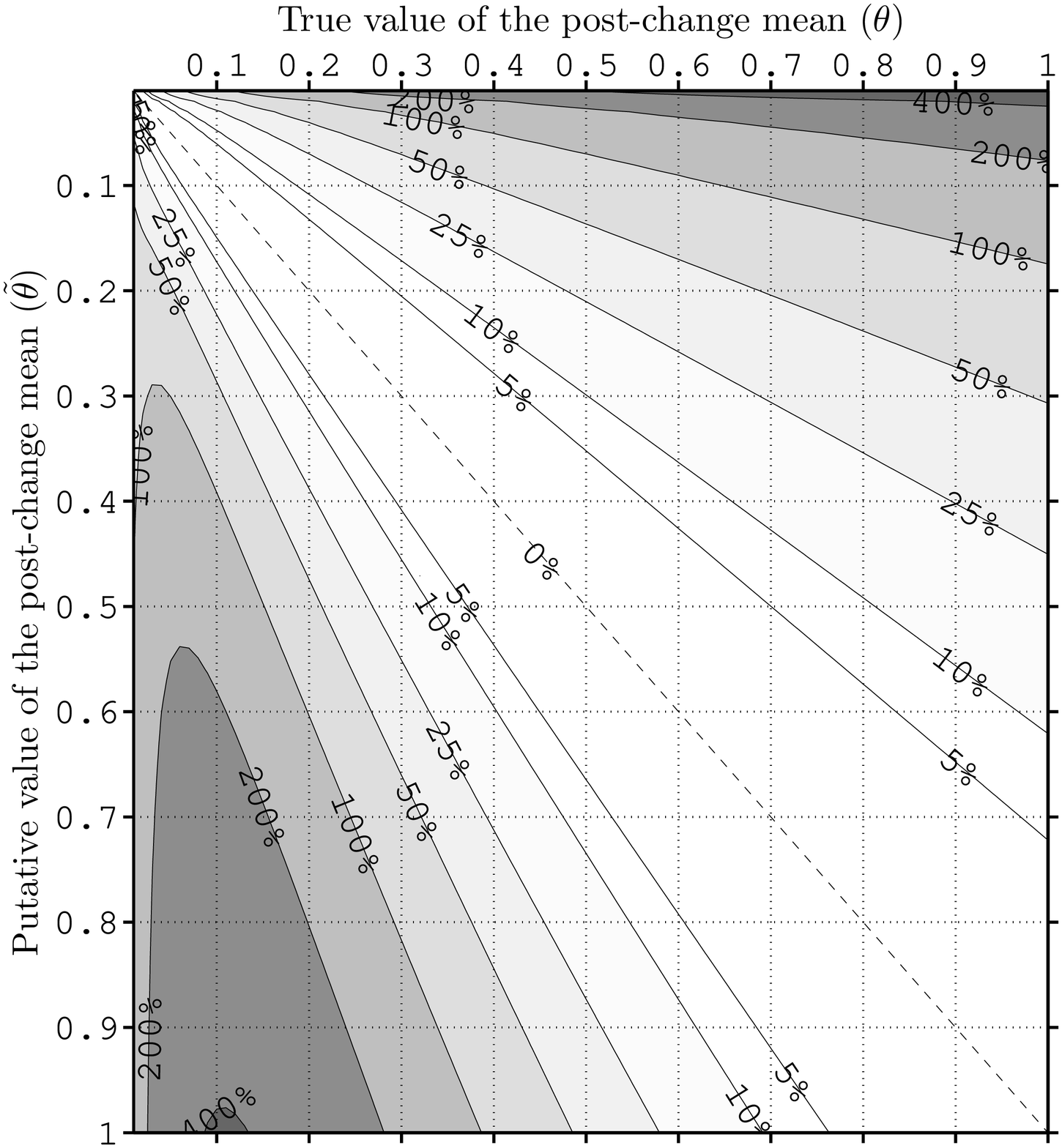}
    \caption{Characterization of $\RE_{\tilde{\theta},\theta}(\mathcal{S}_A)$ as a function of the putative value ($\tilde{\theta}$) and of the true value ($\theta$) of the post-change mean for the ARL to false alarm of $10^4$ (i.e., $\ARL(\mathcal{S}_A^r)=\gamma=10^4$).}
    \label{fig:STADD_r0_vs_theta__gamma1e4}
\end{figure}

Let us now consider the cases when $\gamma=10^3$ and $\gamma=10^4$. Tables~\ref{tab:STADD_r0_vs_theta__gamma1e3} and~\ref{tab:STADD_r0_vs_theta__gamma1e4} and Figures~\ref{fig:STADD_r0_vs_theta__gamma1e3} and~\ref{fig:STADD_r0_vs_theta__gamma1e4} summarize the corresponding results. By and large, the results are similar to the case when $\gamma=10^2$. The difference is that the STADD numbers are higher for higher levels of the ARL to false alarm. This makes perfect sense, because the higher the ARL to false alarm level, the lower the false alarm risk, and the price to pay for being more risk-averse is a higher detection delay. More importantly, Tables~\ref{tab:STADD_r0_vs_theta__gamma1e3} and~\ref{tab:STADD_r0_vs_theta__gamma1e4} and Figures~\ref{fig:STADD_r0_vs_theta__gamma1e3} and~\ref{fig:STADD_r0_vs_theta__gamma1e4} suggest that, as the ARL to false alarm level increases, the SR procedure becomes more sensitive (or less robust) to the error in the post-change parameter. This conclusion can be drawn from the observation that, as the ARL to false alarm level increases, so do the $\RE_{\tilde{\theta},\theta}(\mathcal{S}_A)$ numbers, uniformly across all $\tilde{\theta}$'s and $\theta$'s. This effect of the ARL to false alarm level on the robustness of the SR procedure aligns well with one's intuition.
\begin{table}[!htb]
    \centering
    \caption{Characterization of $\STADD_{\tilde{\theta},\theta}(\mathcal{S}_A)$ and $\RE_{\tilde{\theta},\theta}(\mathcal{S}_A)$ as functions of the putative value ($\tilde{\theta}$) and of the true value ($\theta$) of the post-change mean for the ARL to false alarm of $10^3$ (i.e., $\ARL(\mathcal{S}_A^r)=\gamma=10^3$).}
    \scalebox{0.65}{
    \begin{tabularx}{1.5\textwidth}{@{}l@{\quad}l|*{10}{Y}@{}}
    \toprule
      & & \multicolumn{10}{c}{True value of the post-change mean ($\theta$)}\\[2mm]
      & & 0.1 & 0.2 & 0.3 & 0.4 & 0.5 & 0.6 & 0.7 & 0.8 & 0.9 & 1.0\\
    \cmidrule[0.6pt]{1-12}%
            \multirow{24}{*}{\rotatebox[origin=c]{90}{Putative value of the post-change mean ($\tilde{\theta}$)}}
            &\multirow{2}{*}{0.1}
            &193.5 & 99.38 & 66.23 & 49.63 & 39.7 & 33.09 & 28.38 & 24.86 & 22.12 & 19.94\\
            &
            & $(0\,\%)$ & $(6.92\,\%)$ & $(18.96\,\%)$ & $(32.0\,\%)$ & $(45.13\,\%)$ & $(58.05\,\%)$ & $(70.7\,\%)$ & $(83.03\,\%)$ & $(95.04\,\%)$ & $(106.75\,\%)$\\
\cmidrule{2-12}
            &\multirow{2}{*}{0.2}
            &206.82 & 92.94 & 57.48 & 41.31 & 32.2 & 26.38 & 22.35 & 19.4 & 17.15 & 15.37\\
            &
            & $(6.88\,\%)$ & $(0\,\%)$ & $(3.25\,\%)$ & $(9.87\,\%)$ & $(17.72\,\%)$ & $(26.01\,\%)$ & $(34.44\,\%)$ & $(42.84\,\%)$ & $(51.17\,\%)$ & $(59.37\,\%)$\\
\cmidrule{2-12}
            &\multirow{2}{*}{0.3}
            &231.28 & 96.3 & 55.68 & 38.36 & 29.12 & 23.45 & 19.63 & 16.88 & 14.82 & 13.21\\
            &
            & $(19.52\,\%)$ & $(3.61\,\%)$ & $(0\,\%)$ & $(2.03\,\%)$ & $(6.47\,\%)$ & $(11.99\,\%)$ & $(18.02\,\%)$ & $(24.29\,\%)$ & $(30.64\,\%)$ & $(37.01\,\%)$\\
\cmidrule{2-12}
            &\multirow{2}{*}{0.4}
            &258.75 & 104.22 & 56.96 & 37.6 & 27.74 & 21.91 & 18.09 & 15.41 & 13.43 & 11.91\\
            &
            & $(33.72\,\%)$ & $(12.13\,\%)$ & $(2.3\,\%)$ & $(0\,\%)$ & $(1.41\,\%)$ & $(4.64\,\%)$ & $(8.81\,\%)$ & $(13.47\,\%)$ & $(18.4\,\%)$ & $(23.46\,\%)$\\
\cmidrule{2-12}
            &\multirow{2}{*}{0.5}
            &286.46 & 114.89 & 60.25 & 38.2 & 27.35 & 21.16 & 17.21 & 14.51 & 12.54 & 11.05\\
            &
            & $(48.04\,\%)$ & $(23.62\,\%)$ & $(8.21\,\%)$ & $(1.6\,\%)$ & $(0\,\%)$ & $(1.05\,\%)$ & $(3.52\,\%)$ & $(6.8\,\%)$ & $(10.54\,\%)$ & $(14.55\,\%)$\\
\cmidrule{2-12}
            &\multirow{2}{*}{0.6}
            &313.25 & 127.31 & 65.03 & 39.82 & 27.68 & 20.94 & 16.76 & 13.96 & 11.96 & 10.46\\
            &
            & $(61.88\,\%)$ & $(36.98\,\%)$ & $(16.81\,\%)$ & $(5.9\,\%)$ & $(1.18\,\%)$ & $(0\,\%)$ & $(0.81\,\%)$ & $(2.78\,\%)$ & $(5.43\,\%)$ & $(8.52\,\%)$\\
\cmidrule{2-12}
            &\multirow{2}{*}{0.7}
            &338.67 & 140.82 & 70.98 & 42.26 & 28.57 & 21.13 & 16.63 & 13.67 & 11.6 & 10.07\\
            &
            & $(75.02\,\%)$ & $(51.52\,\%)$ & $(27.48\%)$ & $(12.4\,\%)$ & $(4.44\,\%)$ & $(0.91\,\%)$ & $(0\,\%)$ & $(0.65\,\%)$ & $(2.25\,\%)$ & $(4.46\,\%)$\\
\cmidrule{2-12}
            &\multirow{2}{*}{0.8}
            &362.57 & 154.98 & 77.84 & 45.4 & 29.95 &  21.66  & 16.75 & 13.58 & 11.4 & 9.82\\
            &
            & $(87.37\,\%)$ & $(66.75\,\%)$ & $(39.8\,\%)$ & $(20.77\,\%)$ & $(9.48\,\%)$ & $(3.46\,\%)$ & $(0.73\,\%)$ & $(0\,\%)$ & $(0.53\,\%)$ & $(1.87\,\%)$\\
\cmidrule{2-12}
            &\multirow{2}{*}{0.9}
            &384.98 & 169.47 & 85.43 & 49.16 & 31.76 & 22.5 & 17.09 & 13.66 & 11.34 & 9.69\\
            &
            & $(98.95\,\%)$ & $(82.34\,\%)$ & $(53.44\,\%)$ & $(30.75\,\%)$ & $(16.11\,\%)$ & $(7.47\,\%)$ & $(2.78\,\%)$ & $(0.59\,\%)$ & $(0\,\%)$ & $(0.44\,\%)$\\
\cmidrule{2-12}
            &\multirow{2}{*}{1.0}
            &405.96 & 184.1 & 93.6 & 53.45 & 33.97 & 23.62 & 17.63 & 13.89 & 11.4 & 9.64\\
            &
            & $(109.79\,\%)$ & $(98.08\,\%)$ & $(68.12\,\%)$ & $(42.17\,\%)$ & $(24.18\,\%)$ & $(12.81\,\%)$ & $(6.04\,\%)$ & $(2.28\,\%)$ & $(0.49\,\%)$ & $(0\,\%)$\\
    \bottomrule
    \end{tabularx}
    } 
    \label{tab:STADD_r0_vs_theta__gamma1e3}
\end{table}
\begin{table}[!htp]
    \centering
    \caption{Characterization of $\STADD_{\tilde{\theta},\theta}(\mathcal{S}_A)$ and $\RE_{\tilde{\theta},\theta}(\mathcal{S}_A)$ as functions of the putative value ($\tilde{\theta}$) and of the true value ($\theta$) of the post-change mean for the ARL to false alarm of $10^4$ (i.e., $\ARL(\mathcal{S}_A^r)=\gamma=10^4$).}
    \scalebox{0.65}{
    \begin{tabularx}{1.5\textwidth}{@{}l@{\quad}l|*{10}{Y}@{}}
    \toprule
      & & \multicolumn{10}{c}{True value of the post-change mean ($\theta$)}\\[2mm]
      & & 0.1 & 0.2 & 0.3 & 0.4 & 0.5 & 0.6 & 0.7 & 0.8 & 0.9 & 1.0\\
    \cmidrule[0.6pt]{1-12}
            \multirow{24}{*}{\rotatebox[origin=c]{90}{Putative value of the post-change mean ($\tilde{\theta}$)}}
            &\multirow{2}{*}{0.1}
            &516.46 & 214.84 & 134.68 & 98.04 & 77.09 & 63.53 & 54.04 & 47.02 & 41.63 & 37.36\\
            &
            & $(0\,\%)$ & $(12.39\,\%)$ & $(31.81\,\%)$ & $(51.88\,\%)$ & $(71.7\,\%)$ & $(91.05\,\%)$ & $(109.91\,\%)$ & $(128.3\,\%)$ & $(146.23\,\%)$ & $(163.74\,\%)$\\
\cmidrule{2-12}
            &\multirow{2}{*}{0.2}
            & 606.94 & 191.15 & 107.91 & 74.85 & 57.27 & 46.38 & 38.99 & 33.64 & 29.58 & 26.41\\
            &
            & $(17.52\,\%)$ & $(0\,\%)$ & $(5.61\,\%)$ & $(15.95\,\%)$ & $(27.55\,\%)$ & $(39.49\,\%)$ & $(51.45\,\%)$ & $(63.3\,\%)$ & $(74.98\,\%)$ & $(86.48\,\%)$\\
\cmidrule{2-12}
            &\multirow{2}{*}{0.3}
            & 804.97 & 205.88 & 102.18 & 66.66 & 49.33 & 39.14 & 32.45 & 27.72 & 24.2 & 21.49\\
            &
            & $(55.86\,\%)$ & $(7.7\,\%)$ & $(0\,\%)$ & $(3.27\,\%)$ & $(9.88\,\%)$ & $(17.72\,\%)$ & $(26.05\,\%)$ & $(34.58\,\%)$ & $(43.15\,\%)$ & $(51.7\,\%)$\\
\cmidrule{2-12}
            &\multirow{2}{*}{0.4}
            & 1,054.59 & 244.69 & 106.55 & 64.55 & 45.87 & 35.51 & 28.97 & 24.47 & 21.19 & 18.69\\
            &
            & $(104.2\,\%)$ & $(28.01\,\%)$ & $(4.28\,\%)$ & $(0\,\%)$ & $(2.16\,\%)$ & $(6.79\,\%)$ & $(12.53\,\%)$ & $(18.78\,\%)$ & $(25.3\,\%)$ & $(31.94\,\%)$\\
\cmidrule{2-12}
            &\multirow{2}{*}{0.5}
            & 1,326.24 & 303.4 & 118.7 & 66.3 & 44.9 & 33.76 & 27.03 & 22.53 & 19.33 & 16.93\\
            &
            & $(156.79\,\%)$ & $(58.72\,\%)$ & $(16.17\,\%)$ & $(2.72\,\%)$ & $(0\,\%)$ & $(1.54\,\%)$ & $(4.99\,\%)$ & $(9.39\,\%)$ & $(14.31\,\%)$ & $(19.51\,\%)$\\
\cmidrule{2-12}
            &\multirow{2}{*}{0.6}
            & 1,603.21 & 378.96 & 138.11 & 71.3 & 45.74 & 33.25 & 26.04 & 21.39 & 18.15 & 15.77\\
            &
            & $(210.42\,\%)$ & $(98.25\,\%)$ & $(35.17\,\%)$ & $(10.46\,\%)$ & $(1.88\,\%)$ & $(0\,\%)$ & $(1.15\,\%)$ & $(3.83\,\%)$ & $(7.33\,\%)$ & $(11.32\,\%)$\\
\cmidrule{2-12}
            &\multirow{2}{*}{0.7}
            & 1,876.54 & 468.46 & 164.59 & 79.43 & 48.18 & 33.71 & 25.74 & 20.78 & 17.42 & 15.0\\
            &
            & $(263.35\,\%)$ & $(145.07\,\%)$ & $(61.08\,\%)$ & $(23.05\,\%)$ & $(7.31\,\%)$ & $(1.38\,\%)$ & $(0\,\%)$ & $(0.9\,\%)$ & $(3.04\,\%)$ & $(5.9\,\%)$\\
\cmidrule{2-12}
            &\multirow{2}{*}{0.8}
            & 2,141.43 & 569.14 & 197.93 & 90.76 & 52.17 & 35.05 & 26.02 & 20.6 & 17.03 &  14.51\\
            &
            & $(314.64\,\%)$ & $(197.74\,\%)$ & $(93.71\,\%)$ & $(40.61\,\%)$ & $(16.21\,\%)$ & $(5.4\,\%)$ & $(1.06\,\%)$ & $(0\,\%)$ & $(0.72\,\%)$ & $(2.47\,\%)$\\
\cmidrule{2-12}
            &\multirow{2}{*}{0.9}
            & 2,394.87 & 678.44 & 237.81 & 105.4 & 57.79 & 37.24 & 26.81 & 20.77 & 16.91 & 14.25\\
            &
            & $(363.71\,\%)$ & $(254.92\,\%)$ & $(132.74\,\%)$ & $(63.29\,\%)$ & $(28.71\,\%)$ & $(12.0\,\%)$ &  $(4.15\,\%)$ & $(0.84\,\%)$ & $(0\,\%)$ & $(0.59\,\%)$\\
\cmidrule{2-12}
            &\multirow{2}{*}{1.0}
            & 2,634.79 & 794.01 & 283.75 & 123.41 & 65.11 & 40.34 & 28.12 & 21.28 & 17.02 & 14.16\\
            &
            & $(410.16\,\%)$ & $(315.38\,\%)$ & $(177.7\,\%)$ & $(91.2\,\%)$ & $(45.02\,\%)$ & $(21.31\,\%)$ & $(9.25\,\%)$ & $(3.29\,\%)$ & $(0.68\,\%)$ & $(0\,\%)$\\
    \bottomrule
    \end{tabularx}
    }
    \label{tab:STADD_r0_vs_theta__gamma1e4}
\end{table}

\section{Conclusion}
\label{sec:conclusion}
This work sought to examine what happens to the performance of the Shiryaev--Roberts (SR) procedure when the latter is set up to detect the ``wrong'' change. Specifically, considered within the context of the basic quickest change-point detection problem, the SR procedure was intentionally allowed to be ``out of tune'' in the way of the actual value of a parameter in the observations' post-change distribution (e.g., as a result of the latter known only up to that parameter), and the question was to understand how sensitive Shiryaev's~\citeyearpar{Shiryaev:SMD61,Shiryaev:TPA63} multi-cyclic Stationary Detection Delay (STADD) delivered by the SR procedure is with respect to the severity of the post-change distribution parameter misspecification. To answer this question, we offered a case study where the robustness of the SR procedure was examined numerically in a specific Gaussian scenario. The obtained exhaustive characterization of the SR procedure's robustness can be used to develop the respective theory (which is still missing) and can also provide guidance to practitioners interested in employing the SR procedure. Qualitatively, the overall conclusion of the study was that the less (more) contrast the change and the lower (higher) the false alarm risk, the less (more) robust the SR procedure devised to detect it. This is an expected result.

\section*{Acknowledgements}
The authors are grateful to Dr.~Emmanuel Yashchin of the Mathematical Sciences Department at the IBM Thomas J. Watson Research Center, Yorktown Heights, New York, USA; to Prof.~William H. Woodall of the Statistics Department at the Virginia Polytechnic Institute (Virginia Tech), Blacksburg, Virginia, USA; to Prof.~Sven Knoth of the Department of Mathematics and Statistics at the Helmut Schmidt University, Hamburg, Germany; to Dean Neubauer of Corning Incorporated, Corning, New York, USA; to Prof.~Subha Chakraborti of the Department of Information Systems, Statistics and Management Science at the University of Alabama, Alabama, USA; and to Dr. Ron Kenett of Israel-based KPA Ltd. (\url{www.kpa-group.com}), for the interest in this work and for the constructive feedback that helped improve the quality of the manuscript. Additional thanks goes out to the anonymous referee for the comments and suggestions that helped to ameliorate the paper further.


The effort of A.S.~Polunchenko was supported, in part, by the Simons Foundation (\url{www.simonsfoundation.org}) via a Collaboration Grant in Mathematics (Award \#\,304574) and by the Research Foundation for the State University of New York at Binghamton via an Interdisciplinary Collaboration Grant (Award \#\,66761).

Last but not least, A.S. Polunchenko is also personally thankful to the Office of the Dean of the Harpur College of Arts and Sciences at the State University of New York (SUNY) at Binghamton for the support provided through the Dean's Research Semester Award for Junior Faculty granted for the Fall semester of 2014. The Award allowed to focus on this research more fully.
%

\singlespacing
\bibliographystyle{abbrvnat}      
\bibliography{main,spc}

\end{document}